\documentclass{aa}
\usepackage{graphicx}
\usepackage{txfonts}
\usepackage{rotating}
\usepackage{lscape}
\usepackage{amssymb}
\usepackage{mathtools}

\begin{document}

   \title{Dozens of virtual impactor orbits eliminated \\
          by the EURONEAR VIMP DECam data mining project \\
         }
   \titlerunning{Dozens of VI orbits eliminated by the VIMP DECam data mining project}
      \author{O. Vaduvescu~\inst{1,2,3}
          \and
              L. Curelaru~\inst{4}
          \and
              M. Popescu~\inst{5,2}
          \and
              B. Danila~\inst{6,7}
          \and
              D. Ciobanu~\inst{8}
              }

   \offprints{O. Vaduvescu \\ \email{ovidiu.vaduvescu@gmail.com}}

   \institute{Isaac Newton Group of Telescopes, Apto. 321, E-38700 Santa Cruz de la Palma, Canary Islands, Spain 
%              \email{ovidiu.vaduvescu@gmail.com}
             \and
              Instituto de Astrofisica de Canarias (IAC), C/Via Lactea s/n, 38205, La Laguna, Tenerife, Spain 
             \and
              University of Craiova, Str. A. I. Cuza nr. 13, 200585, Craiova, Romania
             \and
              EURONEAR member, Brasov, Romania 
             \and
              Astronomical Institute of the Romanian Academy, 5 Cutitul de Argint, 040557, Bucharest, Romania 
             \and
              Astronomical Observatory Cluj-Napoca, Romanian Academy, 15 Ciresilor Street, 400487 Cluj-Napoca, Romania
             \and
              Department of Physics, Babes-Bolyai University, Kogalniceanu Street, 400084 Cluj-Napoca, Romania
             \and
              EURONEAR collaborator, Bucharest, Romania 
             }

   \date{Submitted 15 June 2020; Reviewed 23 July 2020; Accepted 27 July 2020}

  \abstract
  % context heading (optional)
  { Massive data mining of image archives observed with large etendue facilities represents a 
    great opportunity for orbital amelioration of poorly known virtual impactor asteroids (VIs).
    There are more than 1000 VIs known today; most of them have very short observed arcs 
    and many are considered lost as they became extremely faint soon after discovery. 
  } 
  % aims heading (mandatory)
  { We aim to improve the orbits of VIs and eliminate their status by data mining the existing 
    image archives. 
   } 
  % methods heading (mandatory)
  { Within the European Near Earth Asteroids Research (EURONEAR) project, we developed the 
    Virtual Impactor search using Mega-Precovery (VIMP) software, 
    which is endowed with a very effective (fast and accurate) algorithm to predict 
    apparitions of candidate pairs for subsequent guided human search. Considering a simple geometric 
    model, the VIMP algorithm searches for any possible intersection in space and time between the 
    positional uncertainty of any VI and the bounding sky projection of any image archive. 
  } 
  % results heading (mandatory)
  { We applied VIMP to mine the data of 451,914 Blanco/DECam images observed between 12 September 2012 
    and 11 July 2019, identifying 212 VIs that possibly fall into 1286 candidate images leading to either 
    precovery or recovery events. 
    Following a careful search of candidate images, we recovered and measured 54 VIs in 183 DECam 
    images. About 4,000 impact orbits were eliminated from both lists, 27 VIs were removed from 
    at least one list, while 14 objects were eliminated from both lists. 
    The faintest detections were around $V$$\sim$24.0, while the majority fall 
    between 21<$V$<23. The minimal orbital intersection distances remains constant for 
    67\% detections, increasing for eight objects and decreasing for ten objects. Most eliminated 
    VIs (70\%) had short initial arcs of less than five days. Some unexpected photometric discovery 
    has emerged regarding the rotation period of 2018~DB, based on the close inspection of longer 
    trailed VIs and the measurement of their fluxes along the trails. 
  } 
  % conclusions heading (optional), leave it empty if necessary 
  { Large etendue imaging archives represent great assets to search for serendipitous encounters 
    of faint asteroids and VIs. 
  } 


   \authorrunning{Vaduvescu et al.}
   \maketitle

%
%________________________________________________________________

% \clearpage

\section{Introduction}

Exploiting the huge potential data wealth of the astronomical image archives has been one of the main 
astrometric aims of the European Near Earth Asteroids Research (EURONEAR) project\footnote{www.euronear.org}, 
besides new observations of fainter 
near Earth asteroids (NEAs), which typically require 2-4~m class telescopes that are difficult to allocate for routine 
observations. Particularly, potential hazardous asteroids (PHAs), virtual impactors (VIs), and imminent 
impactors are in most need of such data for orbital amelioration, which can improve and characterise
their orbits and improve our knowledge of potential future impacts with our planet. \\

Our first such data mining project, Precovery, presented the avenues for the search for NEAs and PHAs 
in a single image archive \citep{vad09}. This tool was applied later by a few teams of students and amateur 
astronomers who searched the entire Canada-France-Hawaii Legacy Survey (CFHTLS) MegaCam archive \citep{vad11}, 
the European Southern Observatory Max Planck Telescope (ESO/MPG) Wide Field Imager (WFI) and the 
Isaac Newton Telescope (INT) Wide Field Camera (WFC) archives \citep{vad13} and the Subaru Suprime-Cam 
archive \citep{vad17}. Together, we improved the orbits of 408 
NEAs based on 425,000 images contained in the respective archives. \\

Our second project, Mega-Precovery, and few related tools have gathered 112 instrument 
archives in the Mega-Archive Structured Query Language (SQL) database, a meta-data collection holding more than 16 million 
images able to search immediately for one given known or unknown moving (Solar System) object or any fixed 
(stellar or extra-galactic) object \citep{vad20}. This big data mining resource is similar to the Canadian 
Solar System Object Image Search (SSOIS)\footnote{http://www.cadc-ccda.hia-iha.nrc-cnrc.gc.ca/en/ssois/.} \citep{gwy12} 
(for moving objects) 
and TELARCHIVE\footnote{http://www.mpe.mpg.de/$\sim$erwin/cod} or other online observatory tools to provide 
searches of major archive collections (only for fixed objects). \\

Little similar data mining work have been conducted by other authors during recent years. 
\citet{sol14} present a citizen-science project implemented online at the Spanish Virtual Observatory 
(SVO) that aims to improve the orbits of NEAs serendipitously falling into the Sloan Digital Sky Survey (SDSS) 
Data Release 8 (DR8) survey (938,046 images). 
This project engaged 3,226 registered users who identified known NEAs and measured their positions, 
improving the orbital elements of 551 NEAs (including only 29 extended arcs). \\

Besides astrometric data mining projects, physical characterisation of NEAs, main belt asteroids 
(MBAs), and other orbital classes can be derived based on major imaging surveys. 
\citet{pop16} conducted the  Moving Objects from VISTA survey (MOVIS) project for the physical characterisation 
of moving objects recovered 
from the Visible and Infrared Survey Telescope for Astronomy (VISTA) VHS-DR3 survey (at the time comprising 
86,562 stack images in four filters). The authors 
derived near-infrared magnitudes and colours of 39,947 objects (including 38,428 MBAs and 52 NEAs, among 
other classes). Again using the MOVIS database, \citet{pop18} provided taxonomic classifications 
of 6,496 asteroids, reporting the albedo distribution for each taxonomic group and new median values 
for the main types. \\

\citet{mah18} present a method to acquire positions, photometry, and proper motion measurements of 
Solar System objects in surveys using dithered image sequences. The authors mined the data from a fraction of 
346 sq. deg of the sky imaged in up to four filters by the VST/OmegaCAM Kilo-Degree survey (KiDS), 
reporting 20,221 candidate asteroids (including 46.6\% unknown objects). \\

\citet{cor19} present two complementary methods to identify asteroids serendipitously observed 
in astronomical surveys. They applied these methods to the 6.4 sq. deg of the sky (30,558 images in 
the J-band) covered by the United Kingdom Infrared Telescope (UKIRT) Wide Field Camera (WFCAM) Transit Survey, 
identifying 15,661 positions of 1,821 asteroids (including 182 potential new discoveries and a few NEA candidates). \\

Since 2002, ESA and ESO have devoted limited Very Large Telescope (VLT) time and some data mining human effort to remove 
VIs \citep{boa03}, and other human resources 
to mine data from some image archives. The Near Earth Objects Coordination Centre (NEOCC) of the European Space Agency (ESA) 
and their collaborating observers targeted about 500 VIs from which almost 100 VIs were removed, a project that include 
the expensive VLT as the key contributor in this project. \\

Recently within EURONEAR we directed our experience towards data mining known VIs, 
due to the non-zero probability of their impact with Earth. To identify candidate images with VIs, one 
needs to take into account the positional uncertainties generated by these poorly observed objects, 
and to this end we designed a new pipeline named the ``Virtual Impactor search using Mega-Precovery'' (VIMP). 
We developed two versions of VIMP software in 2015-2016 and 2018, respectively, testing them with the 
Canada-France-Hawaii Telescope (CFHT) Megacam, then the Blanco DECam archive, whose first findings were 
announced by \citet{vad19a}. \\

The paper is organised as follow. In Sect.~2 we introduce the VI databases and derive some statistics, 
presenting the VIMP algorithm and software. In Sect.~3 we introduce the Blanco/DECam archive, presenting some 
statistics and applying it to the VIMP project and other four large etendue archives. We present the DECam results 
in Sect.~4, followed by conclusions and future projects in Sect.~5. 

%__________________________________________________________________

\section{Virtual impactor data mining}

First, we define virtual impactor asteroids and remind the two existing virtual impactor databases. 
Second, we present the VIMP algorithm and software. 

\subsection{Virtual Impactors}

The modern statistical grounds concerning the recovery and pairing of poorly observed asteroids, 
their multiple orbital solutions, close planetary encounters, and potential risk of collision were 
paved two decades ago by the celestial mechanics group of Andrea Milani at the University of Pisa 
\citep{mil99a,mil99b,mil00,mil09}. Briefly, the region of uncertainty defined by any asteroid is 
characterised by a set of 
multiple solution orbits that acceptably fit the available observations, besides the 
nominal orbit solution, which fits the observations best \citep{mil05a}. 
A virtual asteroid represents a hypothetical minor planet that follows any orbit included 
in the region of uncertainty \citep{mil05b}. 
A virtual impactor (VI) is defined as a virtual asteroid whose orbit could impact Earth or 
another planet sometime in the future (although with a very small probability), according to 
current observations \citep{mil05}. \\

\subsection{Virtual impactor databases}

More than 22,600 NEAs are known today (April 2020)\footnote{https://www.minorplanetcenter.net/}
thanks to the major U.S. lead surveys, from which about one thousand are VIs, and this number has 
grown by about 100 VIs yearly in the last few years. Aiming for completion during the 
VIMP project, we joined the two available VI databases. 

\subsubsection{CLOMON at NEODyS/Pisa}

The Near-Earth Object Dynamics Site (NEODyS) is an online information 
service\footnote{https://newton.spacedys.com/neodys/} for NEAs developed in the late 1990s 
by the A. Milani group at the University of Pisa \citep{che99}. 
NEODyS updates its NEA database daily, comprising observations available from the Minor 
Planet Centre (MPC). It adjusts orbital elements, runs ephemerides with positional 
uncertainties, and checks close encounters with planets using the powerful ORBFIT 
software package\footnote{http://adams.dm.unipi.it/orbfit/} previously developed by 
the same team. \\

CLOMON is a fully automated system that has monitored all known NEAs for potential collisions 
with Earth since November 1999, in conjunction with NEODyS and ORBFIT \citep{che00}. 
This system maintains a prioritised queue of objects to be processed, based on a 
score that attempts to quantify the likelihood of finding any collision solution. 
For each object at the head of the queue, CLOMON conducts daily searches for virtual orbits that are
compatible with the available observations and that could lead to virtual impacts with Earth within 
the next 100 years \citep{che00}. The resulting asteroids that have at least one virtual 
impact are posted on the NEODyS risk list. 
In 2002, CLOMON was replaced by the second generation CLOMON2 impact monitoring system, 
which uses the line of variations (LOV) approach, sampling the LOV one-dimensional 
subspace to perform the sampling of the six-dimensional confidence region 
\citep{mil09,tom05}. 
By 4 April 2020, there were 1039 objects on the NEODyS/CLOMON2 risk list, including 
VIs that have at least one virtual impact with Earth in the next 100 years. 

\subsubsection{Sentry at JPL/CNEOS}

In 2002, Jet Propulsion Laboratory (JPL) released the 
Sentry\footnote{https://cneos.jpl.nasa.gov/sentry/} asteroid impact monitoring 
system, now hosted under the JPL Center for NEO Studies (CNEOS). Sentry is very similar 
to CLOMON of NEODyS, predicting future close Earth approaches along with their associated 
impact probabilities and cross-checking results for objects of highest risk \citep{yeo02}. 
The asteroid orbits, close approaches, and impact predictions are revised daily by Sentry, 
which updates its risk table, quantifying the risk posed by the tabulated objects using 
both the Torino Scale (designed primarily for public communication of the impact risk) 
and the Palermo Scale (designed for technical comparisons of the impact risk). 
By 4 April 2020 there were 1004 VIs on the Sentry risk table. We use this database to 
derive some statistics, making use of the Sentry API interface (available since 2016) 
to retrieve the currently observed arcs and diameters. \\

Figure~1 shows the histogram counting all known VIs as a function of their observed arcs. 
Most VIs have very few observations, the great majority (97\%) having arcs less than 47 days. 
Three objects have arcs less than one hour, namely: 2006~SF281, 2008~UM1, and 2008~EK68, all 
of which were discovered and observed by G96 Catalina Mt. Lemmon survey only. 
Eighty-four VIs currently have arcs shorter than ten hours, and 165 objects (16\%) have arcs observed 
for less than one day. 
A bit more than a quarter (28\%) of currently known VIs have arcs of less than 2 days, 
about half (48\%) have been observed for less than 5 days, while the other half (50\%) have 
arcs between 5 and 100 days. 
Only eight VIs have arcs longer than one year, namely 2011~TO, 2014~KS76, 2000~SG344, 443104, 
(99942) Apophis, (410777), (101955) Bennu, and (29075), which has been known for 66 years. \\

\begin{figure}
\centering
\includegraphics[angle=0,width=9cm]{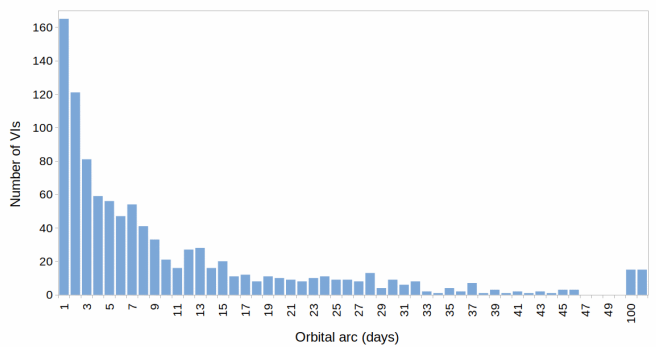}
\begin{center}
\vspace{-0.2cm}
\caption{Histogram counting 1004 VIs from the Sentry database (4 Apr 2020) function of their current observed arc.
         The last two bins are much larger, to fit the best observed objects on the same plot. } 
\label{fig1}
\end{center}
\end{figure}

Figure~2 shows the histogram counting all known VIs function of their diameters, 
assuming uniform spherical bodies with a visual albedo of 0.154 in accordance with the 
Palermo Scale. 
The great majority (949 objects or about 95\%) have diameters smaller than 100~m,
but 54 objects are larger than 100~m, the largest being (29075) of about 1.3 km,
2014~MR26 (about 900~m), 2001~VB, 2001~CA21, 1979~XB (about 700~m), 2014~MV67, 
2015~ME131, (101955) Bennu, 2017~SH33, and 2016~WN55 (about 500~m). 
About three quarters (771 objects or 77\%) are smaller than 30~m, 
and about one third (356 objects or 35\%) are smaller than 10~m. 

\begin{figure}
\centering
\includegraphics[angle=0,width=9cm]{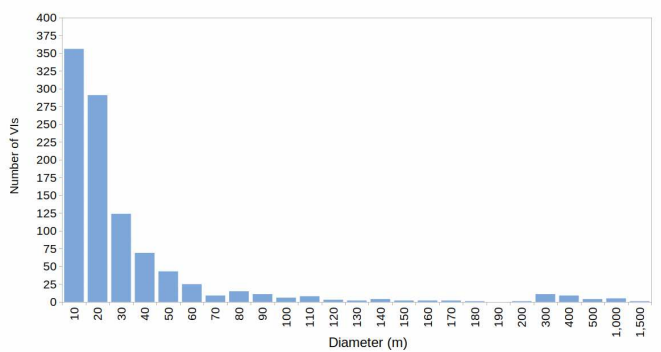}
\begin{center}
\vspace{-0.2cm}
\caption{Histogram counting 1004 VIs from the Sentry database (4 Apr 2020) function of their diameter, 
         assuming 0.154 albedos and spherical bodies. 
         The last five bins are larger (above 200~m), to fit the largest 30 objects. } 
\label{fig2}
\end{center}
\end{figure}

\subsection{VIMP software}

We present next the algorithm of search and the selection criteria of image candidates possible to hold VIs. 

\subsubsection{Algorithm}

Figure~3 presents the flowchart of the VIMP algorithm and databases. \\

\begin{figure}
\centering
\vspace{-0.4cm}
\includegraphics[angle=0,width=9cm]{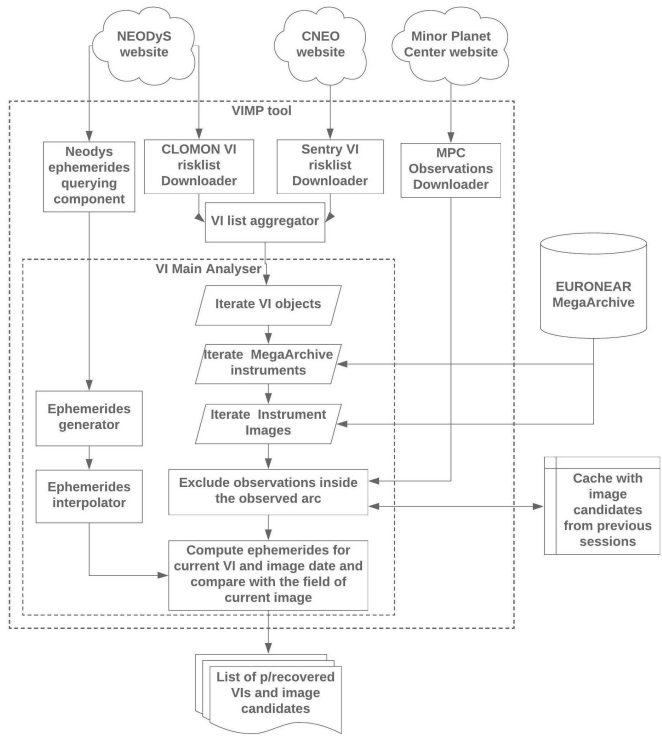}
\begin{center}
\caption{Flowchart presenting the VIMP software algorithm integration. } 
\vspace{-0.2cm}
\label{fig3}
\end{center}
\end{figure}

The EURONEAR Mega-Archive meta-database (Blanco/DECam or another available instrument) 
feeds VIMP with images to be queried. The NEODyS/CLOMON and JPL/Sentry risk list websites 
are called once, being amalgamated into a common VI list, while the Minor Planet Center (MPC) 
observations database provides the observed date limits for each VI. \\

As an alternative to data mining all VIs, VIMP can accept as input a list of VIs or only one 
VI. This option is very useful for second or further VIMP runs (referred to as VIMP2, 3, 4, and so on,   
in our work and run for a few objects) that aim to remove VIs that could not be removed 
after the initial VIMP findings. 
Typically one day following each submission, the improved NEODyS orbit (adding VIMP data) 
becomes available, then a new VIMP2 search becomes more efficient, as uncertainty regions 
shrink. We have only run VIMP2 searches for objects not removed as VIs after the first VIMP 
phase, finding  new apparitions of three objects in VIMP2: 2014~HT197 (removed), 2016~JT38 
(PHA recovered on two nights after VIMP2 but not removed yet), and 2014~HD199 (removed after 
two more nights recovery). For the great majority of cases, VIMP2 provided the same results 
as VIMP1. \\

Probably only some marginal improvement of the orbit determination can be obtained by 
adding the astrometric position inside the observed arc, although there might be cases 
when more substantial improvement could be archived (for example when the nights at the 
start and end of observations are more distant in time). As a consequence, in this project 
we decided only to mine the data of images taken outside the observed arc, bounded by two 
equal time intervals before the discovery or after the  last observation of any given object. 
During this work we refer to these two intervals commonly as the ``arcbound''. Taking into 
account the typical fast growing positional uncertainties in time for poorly observed VIs, 
we limited the arcbound to 60 days before and after the observed arc, but this value could 
eventually be enlarged in other VIMP runs (running the risk of an increase in the number of false 
candidates and the execution time). 
The election of 60 days was decided based on the limited human resources available 
for the actual search in this demonstrator project. Based on our experience with the size
of the uncertainty ellipses, beyond this limit the positional uncertainty grows with time 
excessively, increasing the probability for targets to actually fall into many CCDs, into a 
gap between CCDs, or even outside the field, which slows down the efficiency of the whole 
project. \\

The VIMP algorithm can be described by the following steps: \\

\begin{enumerate}

\item The CLOMON and Sentry databases (which update regularly) are queried by VIMP
and amalgamated into the list of VIs known to date. 

\item First, the VI list is iterated, setting one given object to be searched in the  
instrument archive (in this case Blanco/DECam) or a few selected archives (possible to 
cycle in the same VIMP run). 

\item The MPC observations database is queried for the given VI, providing the first 
and last observing dates necessary to provide the arcbound for subsequent data mining. 

\item Giving the resulting arcbound, the NEODyS ephemeride is queried for the given 
VI and observatory (the MPC code being taken from the archive), using a step of one day, caching 
the results to use later for all other images observed in the same arcbound. 

\item Second, all instrument archive images are iterated, reading the observing date 
and telescope pointing coinciding with camera centres ($\alpha_c$, $\delta_c$), in order
to select the images falling within the arcbound and close enough to the target. 

\item For any selected image, the VI ephemeride is interpolated linearly to 
provide approximate positions and uncertainties based on the image observing date. 

\item If the predicted magnitude is below the instrument limiting magnitude (given
in the input ($V=26$ for DECam), then the search continues. 

\item If the object moves more slowly than $\mu<2$ deg/day, then VIMP interpolates the sky 
position given the available ephemeris grid. 

\item If the object moves more quickly than the above limit, then VIMP sends a new NEODyS 
ephemeride query for the arcbound using a one hour time step to refine predictions for 
cases of close approaches when the ephemeris becomes non-linear in the previous one 
day step. 

\item For any of the above cases, the software calculates the  apparent separation on the sky 
(in arcsec) between the telescope pointing (assumed to coincide with the image centre 
allowing a small uncertainty bounding set in the input) and the interpolated position 
of the object, comparing this separation with the sum of the half-diagonal camera 
(known in arcsec) plus the current positional uncertainty (the NEODyS ellipse major axis, 
assumed at this step to be free in orientation). 

\item If the calculated separation is smaller than the second above quantity, then 
the object and image become a possible candidate pair (asteroid, image) for detailed 
analysis in the next step. 

\item For each possible candidate pair, VIMP again queries NEODyS to obtain the accurate 
ephemeride for the actual image date and time together with its associated uncertainty position 
ellipse (the major and minor axis, plus the sky position angle of the major axis), saving this data
and building the region file to be used in the manual search process. 

\item The actual search area for the actual candidate pair corresponds to the common 
area defined by the intersection of the $3\sigma$ ellipse uncertainty region with the 
rectangular image camera field. Therefore, VIMP needs to solve a geometrical problem 
involving the equation of a tilted ellipse and the equations of the sides of a rectangle. 

\end{enumerate}

Between one and eight solutions will be solved by VIMP based on the quadratic ellipse 
equation to provide possible intersections between the rectangle and the ellipse. 
Based on the ratio of the ellipse section inside and outside the rectangle, a probability 
percentage is recorded in order to prioritise the manual search. \\

Next we  detail the mathematics used to solve the intersection problem. Taking into account
the right ascension ($\alpha$) and declination ($\delta$) of the target, the equation of 
the uncertainty region ellipse is 

\begin{equation}
\;\;\;\;\;\; 
\frac{( \Delta\alpha \cos\theta + \Delta\delta \sin\theta )^2}{a^2} + \frac{( \Delta\alpha \sin\theta - \Delta\delta \cos\theta )^2}{b^2} = 1 
,\end{equation}
\noindent
with $\Delta\alpha = (\alpha-\alpha_c)$, $\Delta\delta = (\delta-\delta_c)$, 
where $(\alpha_c,\delta_c)$ is the telescope pointing (centre of the image), 
$\theta$ is the position angle, and (a,b)is  the semi-major axis of the uncertainty 
region ellipse (provided by NEODyS). \\

We assume a rectangular camera image (valid for all CCDs and most mosaic cameras), 
otherwise we define a rectangle bounding the mosaic camera (for example DECam whose shape is 
circular like). Also, we assume the camera to be mounted parallel to the celestial equator, 
noting the image bounding limits by $(\alpha_{min},\alpha_{max})$ (in $\alpha$) and 
$(\delta_{min},\delta_{max})$ (in $\delta$) equatorial directions. \\

Projected on a plane tangent to the sky sphere, the elliptical uncertainty region
associated with the nominal VI predicted position might intersect or not the rectangle 
bounding the camera image. Any possible solution $(\alpha,\delta)$ for any potential 
intersection between the two shapes must satisfy both the equation of the ellipse 
and the equation of the bounding of the rectangle. \\

Equation (1) can be transformed into a quadratic polynomial equation with two variables, 
$\Delta\alpha$ and $\Delta\delta$: 

\begin{equation}
\begin{multlined} 
(\Delta\alpha)^2(a^2\sin^2\theta+b^2\cos^2\theta) \; + \\
+ 2(b^2-a^2)\Delta\alpha\Delta\delta\sin\theta\cos\theta \; + \\
+ (\Delta\delta)^2(a^2\cos^2\theta+b^2\sin^2\theta)-a^2b^2 = 0.
\end{multlined}
\end{equation}

First, fixing $\Delta\alpha = (\alpha_{min}-\alpha_c)$ (which define the left-hand border 
of the image), Eq. (2) will provide a maximum of two solutions for $\delta$. Similarly, 
$\Delta\alpha = (\alpha_{max}-\alpha_c)$ (right-hand border of the image) will provide 
another possible two solutions for $\delta$. 
Second, fixing $\Delta\delta = (\delta_{min}-\delta_c)$ (bottom side of the image) will 
provides a maximum of two solutions for $\alpha$, while $\Delta\delta = (\delta_{max}-\delta_c)$ 
(upper side of the image) will provide another maximum of two solutions for $\alpha$. \\

Overall, a maximum of eight solutions are possible. A minimum of two solutions mean some arcs of 
the ellipse cross the rectangle, thus the VI could fall inside the image, so in this 
case VIMP outputs a candidate pair. 
In case the ellipse does not intersect the rectangle, then it is located either 
entirely outside the rectangle (the VI is outside the image) or entirely inside the 
field (the target falls inside the image). In this case, VIMP outputs the candidate pair. 

\subsubsection{Selection criteria}

To output candidate images of the visible VIs falling within the field of DECam, we used the 
following four selection criteria: 

\begin{itemize}

\item[$1)$] The positional ellipse uncertainty region needs to be included or to intersect 
the rectangle defining the limits of the camera (disregarding the missing corners). 

\item[$2)$] To greatly improve the existing orbit, we only searched outside the observed 
arcs, disregarding images dated inside the actual orbital observed arc of any given VI. 
Past data mining projects showed very little evidence for orbital improvement by adding 
new data inside the observed arcs \citep{vad11,vad13}. 

\item[$3)$] To minimise false candidates due to a drop in brightness and growing positional 
uncertainty, we only searched 60 days after or before the last or first observation 
available. Given the fact that the great majority of VIs are discovered during Earth 
fly-bys, this limit was set to accommodate brighter apparitions and smaller uncertainties 
in positions that are growing outside the observer arc. Setting such a limit in time also 
helps to reduce significantly the VIMP running time. 

\item[$4)$] To avoid the risk of losing fainter objects, we set a limiting magnitude of $V=26$, 
chosen to accommodate the best sky conditions for the 4.2~m Blanco telescope (about $V$=24.5 
using 1 min exposures are able to provide star-like images for best the signal-to-noise ratio (S/N)
of slower moving VIs)\footnote{Using the S/N Excel sheet calculator \texttt{DECam\_ETC-ARW7.xls} at 
\texttt{http://www.ctio.noao.edu/noao/node/5826} and comparing with S/N calculators 
for other 4-m class telescopes.}, any eventual errors in reported magnitudes used to calculate 
$H$ and expected apparent magnitude (about 0.5 mag), and any major jump by a maximum of 
$1.0$ mag (known to exist) in the light curves of very small asteroids. 

\end{itemize}

No criteria were based on exposure time, filter, number of encounters in the same region or night, 
Moon phase, and separation or target altitude. 

%______________________________________________________________

\section{Applications}

In summer 2018 we tested VIMP, first with the ESO 2.5~m VST/OmegaCam instrument, which 
produced some false candidate encounters, although we identified and reported  the finding of
2014~UX34 inside the observed arc. Next, we decided to increase the chances of finding candidates,
so we approached larger facilities. 

\subsection{Blanco/DECam archive}

Here we introduce the DECam camera, the image archive and surveys. 

\subsubsection{DECam camera}

The 4.0~m diameter Victor Blanco telescope was commissioned in 1974, and is located at 2207~m 
altitude at the Cerro Tololo Inter-American Observatory (CTIO) in Chile, which is part of the National 
Optical Astronomy Observatory (NOAO) of the United States. Mounted at the prime focus of Blanco (instrument ratio F/2.7), in Sep 2012 the 
Dark Energy Camera (DECam) became the most powerful survey imaging instrument 
in the world\footnote{Being surpassed since August 2013 only by Subaru/HSC.}, 
with an etendue of $32.7$ sq.m $\times$ sq. deg\footnote{Based on the collecting mirror surface 11.8 sq.m and the covered sky of DECam 
2.78 sq. deg.}. 
Endowed with 62 $2048 \times 4096$ pixel, red-sensitive science CCDs\footnote{Plus 12 smaller 
2K $\times$ 2K CCDs for guiding, focus, and alignment.}, 
on the sky DECam covers almost 3 square degrees in a hexagon inscribed within a 2.2 degrees 
diameter field at a resolution of $0.263^{\prime\prime}$/pixels \citep{hon08}. \\

\subsubsection{DES and DECam surveys}

During 758 nights observed between 31 Aug 2013 and 9 Jan 2019, the Dark Energy Survey 
(DES)\footnote{https://www.darkenergysurvey.org} has imaged 5000 sq. degrees of the southern 
sky using DECam in five bands ($grizY$), with the main goal of probing dark energy and 
testing alternative models of gravity by studying large-scale structures, galaxy cluster 
counts, weak gravitational lensing, and supernovae. 
There were 510 science papers mentioned in the DES publications page by 15 April 2020. 
Besides the main cosmology goals of DES, other topics can benefit from the legacy prospects 
of DES and DECam, namely: the Solar System, stellar studies, galactic studies, globular clusters, local 
extra-galactic studies, gravitational waves, galaxy evolution, clusters, gravitational lensing, 
quasars, supernovae, and other transient events \citep{abb16}. \\

Besides DES, the vast majority of DECam time has been devoted to other ``non-time-domain'' 
imaging surveys, such as VST-ATLAS, DECaLS, DECaPS, DeROSITAS, BLISS, MagLiteS \citep{daw19}, 
adding other community science based on regular time allocation. As part of DECam Solar System 
science and other related technical achievements, the following topics are notable. 

The NEO DECam survey lead by Lori Allen \citep{all15} lasted
30 nights covering 975 sq. deg. It aimed to discover NEOs using a tailored DECam moving object 
detection system (MODS) \citep{val15} with the intention of characterising the distribution of small 
10-100~m NEOs \citep{tri17}. 
\cite{whi19} developed and tested a new Graphics Processing Unit (GPU) kernel-based computational technique, 
aiming to detect slow moving asteroids based on 150 sq. deg observed in the DECam High Cadence 
Transient Survey (HiTS) in which the authors discovered 39 Kuiper belt objects. 
\cite{fue14} measured the size distribution of small Centaur asteroids (1-10 km) and their 
evolutionary links to TNOs and Jupiter family comets. 
Two Neptune L4 Trojan asteroids in the DES-supernova survey were discovered \citep{ger16}. 
A new method for estimating the absolute magnitude frequency distribution of NEAs, 
based on observations around opposition for a sample of 13,466 NEAs was developed, which includes the smaller 
but deeper DECam NEO Survey \citep{val18,val19}. 
\cite{mar20} undertook a deep search (limiting magnitude $V\sim25$) of Earth Trojans in 24 sq. degrees
around the L5 point. It did not discover any such new object, placing constraints on the size of such a 
possible population.

\subsubsection{DECam image archive}

By 11 July 2019, the public DECam archive included 451,914 images observed between 
12 September 2012 and 11 July 2019. 
Between 60 and 70 thousands DECam images were observed in a typical year, namely: 
2012 - 15036 (3\% mostly for commissioning), 2013 - 60815 (13\%), 2014 - 75347 (17\%), 
2015 - 69207 (15\%), 2016 - 61560 (14\%), 2017 - 64264 (14\%), 2018 - 64226 (14\%), and the 
first half of 2019 - 41465 (9\%). \\

Figure~4 plots all DECam pointings (cyan dots) in a rectangular projection. 
The p/recovered\footnote{Either recovery or precovery events (resulting from images 
taken before discovery).} are overlayed with larger symbols. \\

\begin{figure}
\centering
\includegraphics[angle=0,width=9cm]{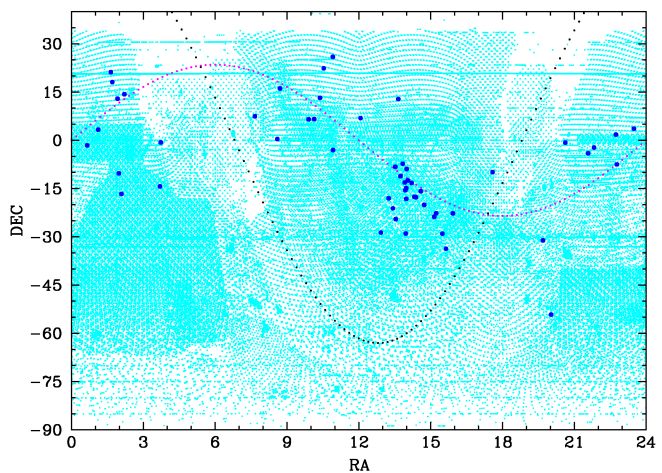}
\begin{center}
\caption{A total of 451,914 science pointings observed by DECam between 12 September 2012 and 11 July 2019 (cyan points). 
         The precovered or recovered VIs are plotted with larger blue symbols, the ecliptic is traced with magenta, and 
         the galactic plane with black points. } 
\label{fig4}
\end{center}
\end{figure}

All DECam images can be counted based on filters as follows: $r$ - 103152 (23\% from all images), 
$g$ - 101758 (22\%), $z$ - 85250 (19\%), $i$ - 72897 (16\%), $Y$ - 36429 (8\%), $VR$ - 30972 (7\%), 
$u$ - 16971 (4\%) and no filters 4485 (1\%). Figure~5 counts all 451,914 DECam images as a 
function of exposure times. The majority (81\%) were 
exposed for less than 120~s, and a big fraction (25\%) were exposed for around 90~s, which makes the data mining 
of the DECam archive a good opportunity for NEAs and other Solar System research. 

\begin{figure}
\centering
\includegraphics[angle=0,width=9cm]{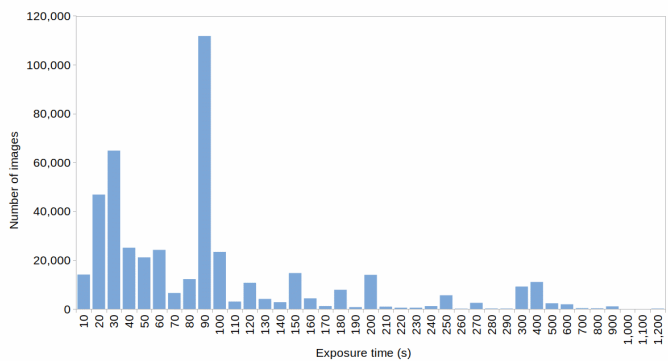}
\begin{center}
\caption{Histogram counting exposure times for 451,914 science pointings observed with DECam between 
         12 September 2012 and 11 July 2019. Please note the change in bins above 300~s, to be able to accommodate 
         longest exposed images. } 
\label{fig5}
\end{center}
\end{figure}

\subsection{Other archives}

We considered a small number of other facilities with large etendues, namely: CFHT-MegaCam, Pan-STARRS1, Subaru/HSC,
and VISTA/VIRCAM. We ran VIMP for about 950 VIs known at 25 Sep 2019 and we obtained the VI candidates 
to be p/recovered in these archives and the number of image candidates. We include the results 
in Table~1, listing the telescope, camera, number of CCDs, querying limiting magnitude $V$, effective 
field of view (FOV) in square degrees, the etendue $A\Omega$ (quoting our DECam number and \cite{die11} 
for other instruments), number of queried images, observing date interval, number of VIs, and image 
candidates to be p/recovered. One can observe the high number of candidates for DECam, due to the setup 
of a very optimistic limiting magnitude ($V=26$). 

%______________________________________________________________

\section{DECam Results}

Next we present the results of search for VIs in the DECam image archive. 

\subsection{Image candidates}

In Sect. 4.3 we list the selection criteria used to query the DECam archive. 
Running VIMP for 451,914 DECam images observed between 12 September 2012 and 11 July 2019 
for about 920 VIs known by 11 July 2019 using criteria exposed in Sect.~2.3.2, 
we obtained 212 VI candidates and 1286 image candidates for p/recovery. \\

Many candidate objects were actually discovered by DECam (marked by Obs. W84 in 
the 'Disc' column of Table~2) and could be linked to the DECam NEO Survey programme (PI: 
Lori Allen), which observed many opposition fields over 30 nights (three runs in 
2014, 2015, and 2016) using  a $VR$ filter with exposures of 40~s \citep{all14}. 
We could p/recover 21 such VIs (39\% of total) from which 20 objects appeared 
in images of their runs, and we removed 16 VIs (including 9 from both lists). 

\subsection{Search procedure}

Most searches were performed by one main reducer, helped by another reducer 
during the second half of the project, while another person checked the results and orbital fits throughout 
the whole project. When we encountered a target, measured its positions, and tested the orbital fit, 
we then submitted the report to the Minor Planet Centre (MPC), which published the results in MPS circulars 
included in the last column of Table~2. \\

Given one target (for example VI 2014~HN197), we considered all candidate images ordered in time that were predicted 
by VIMP to hold the target. 
Given each candidate image, we searched the NOAO Science 
Archive\footnote{http://archive1.dm.noao.edu/search/query} (ex \texttt{c4d\_140423\_022416} selecting 
Telescope \& Instrument = CTIO 4m + DECam imager and no other constraint), using mostly calibrated 
images (extension \texttt{\_ooi}) or raw images when the first were not available (\texttt{\_ori}). 
% OBS Luci: resampled images not used (extension \texttt{_opi}}
Then we downloaded the resulting file (about 300 MB each DECam image archived in NASA funzip format 
\texttt{fz} - ex \texttt{c4d\_140423\_022416\_ooi\_VR\_v1.fits.fz}). 
Given the pair of VIMP output candidate images closest in time and preferably taken with the same 
filter and exposure, we used the first as an input in the EURONEAR FindCCD 
tool\footnote{http://www.euronear.org/tools/findccdaster.php} (for example \texttt{c4d\_140423\_022416}, $PA=0$, 
and $Instrument$ Blanco/DECam) to predict the candidate CCD holding the target (VI or any other NEA), 
overlaying the NEODyS positional uncertainty ellipse (in red) above the DECam mosaic geometry of DECam. 
Most ellipses were relatively small (fitting inside one CCD), but others were larger and could cover 
two or more neighbouring CCDs that then needed to be  searched (Fig.~6). \\

\begin{figure}
\centering
\vspace{0.5cm}
\fbox{ \includegraphics[angle=0,width=7cm]{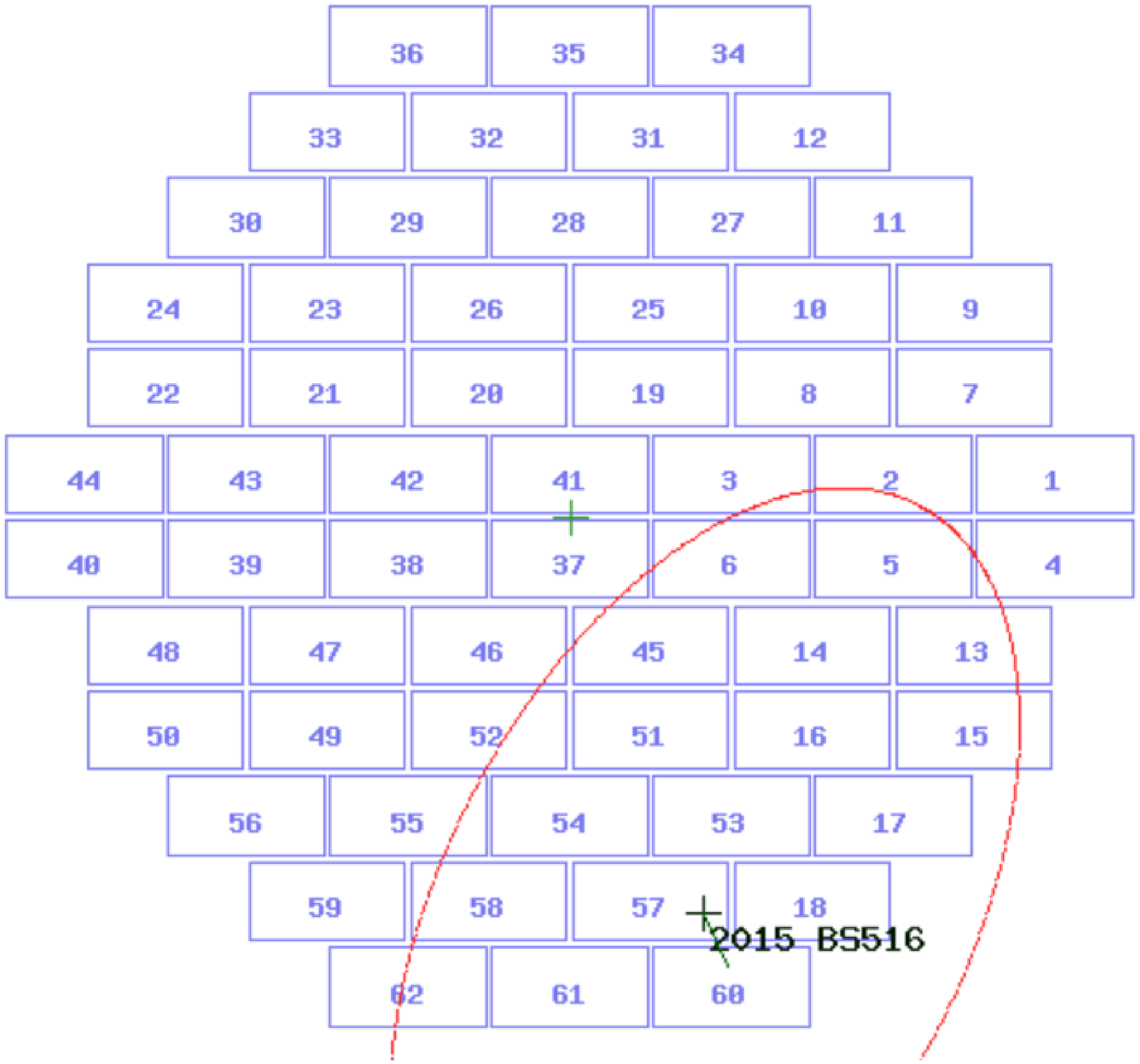} }
\begin{center}
\caption{FindCCD plot showing the Blanco-DECam overlay of the image \texttt{c4d150213085354ooigv1} 
         and the uncertainty position as of Jan 2019 of the poorly observed NEA 2015~BS516 (bordered in 
         red) covering many CCDs that needed careful searching by a team of three people. } 
\label{fig6}
\end{center}
\end{figure}

We performed the search under Windows 10, automating the work for many files in batch scripts. 
To unpack the \texttt{fz} images, we used the NASA Funpack 
software\footnote{https://heasarc.nasa.gov/fitsio/fpack}. 
To extract CCD planes from the resulting MEF image, we used the Aladin desktop 
software\footnote{https://aladin.u-strasbg.fr/AladinDesktop}. 
Another alternative could be to use SAOImage DS9
software\footnote{https://sites.google.com/cfa.harvard.edu/saoimageds9} , which can directly 
open the \texttt{fz} packed multiple extension cube images, load the candidate CCD slice, then save it 
as FITS. 
For the actual image search, we used Astrometrica\footnote{http://www.astrometrica.at} 
to load the CCD slices of the same candidate region to hold the target, checking the CCD centres 
($\alpha,\delta$) versus the table output of FindCCD. 
Targets presenting small positional uncertainties were searched around their NEODyS predicted 
ephemerides, while those with larger uncertainties or/and falling in dense stellar regions were 
searched inside the uncertainty ellipses (with semi-major axis and position angles taken from 
NEODyS) overlaid in DS9 software as Region 
files\footnote{According to the Section IX. A method to plot 
the line of variation from http://www.euronear.org/manuals/Astrometrica-UsersGuide-EURONEAR.pdf.}
output of VIMP (Fig.~7). 

\begin{figure}
\centering
\includegraphics[angle=0,width=8cm]{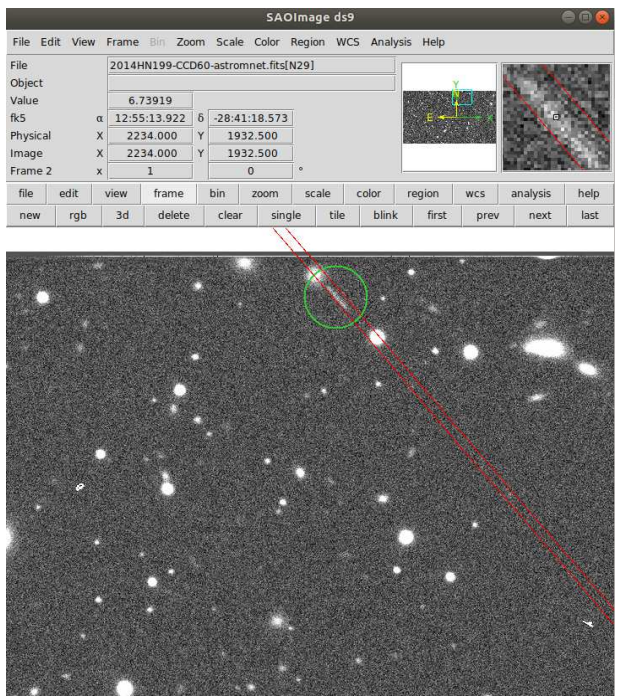}
\begin{center}
\caption{DS9 display search of the DECam \texttt{c4d\_140430\_232945\_opi\_VR\_v1} CCD~60 image, overlaying the 
         NEODyS uncertainty region (bordered in red) of the trailed VI 2014~HN199 (marked in blue). } 
\label{fig7}
\end{center}
\end{figure}

\subsection{False candidates}

About three quarters of VI candidates could not be found in candidate images due to many factors not taken 
into account in our selection criteria, which we recall here. \\

The real shape of DECam is actually hexagon-like, which we approximated by a rectangle, 
thus some VIs could be missed in the corners of the rectangle. Taking into account the DECam geometry 
including the gaps, we calculated a 26\% loss in missing corners, which statistically means a similar 
percentage of lost VIs. \\

The small gaps between CCDs could cause a few VIs to fall inside and be lost. 
Taking into account the widths of all gaps\footnote{Based on http://www.ctio.noao.edu/noao/node/2250} 
and counting the entire area covered by gaps and that of all 62 CCDs, we calculated a DECam filling factor 
of $87.4\%$. This means that for about one in ten encounters, the VI could escape into such a gap, and 
in our project we encountered three such cases, namely: 
2017~WZ12 (a long trail, one end in a gap, the other end above a star), 
2015~XP (VIMP2 re-run of VIMP, after orbital improvement by VIMP1, aiming to eliminate all impact orbits), 
and 2015~HE183 (this object was precovered based on four images, with a fifth image not reported due to one end 
of the trailing object having fallen into a gap). \\

After scrutinising some uncertain objects by searching carefully inside long ellipse 
uncertainty regions, some VIs did not show up, due to the fact that the majority of uncertainty 
regions fell outside the DECam field. One such very difficult case was 2015~BS516 ($V=22$ and long 
$23^{\prime\prime}$ trail), which required the increased effort of three people who carefully searched 
over 24 CCDs using both blink and track and stack. Figure~6 plots the circumstances of this candidate,
whose search was motivated by its large size ($H=19.2$, diameter 490~m assuming albedo 0.154). Its 
uncertainty ellipse covered many CCDs on the DECam candidate image (Fig.~6). In the 
meantime, F51 precovered the object, then G96 and other stations recovered it in 2019, which 
removed this VI. Based on the new orbit, we confirmed that the target fell into the defective CCD number 61. \\

Two CCDs ceased to work\footnote{http://www.ctio.noao.edu/noao/node/14946} 
(CCD~2 = S30 between Nov 2013 and Dec 2016, CCD~61 = N30 after Nov 2012) and we encountered two 
such cases: 2014~EU VIMP2 second precovery attempt following first precovery VIMP1, and 
2015~BS516 see above). \\

Twenty-six candidate images possibly holding five VIs (2013~CY, 2013~GM3, 2013~HT150, 2013~RZ53, 
and 2013~TP4) were affected by a NVO problem leading to bad declination for some images close to 
$\delta=0.$ This affected our DECam archive indexing and therefore VIMP predictions\footnote{Apparently 
this problem was solved in the new NVO NOIRLab Astro Data Archive interface.}. \\

The accuracy of Blanco pointing is typically good within $10-20^{\prime\prime}$ 
across the sky\footnote{http://www.ctio.noao.edu/noao/content/note-blanco-pointing}. To accommodate 
possibly larger pointing errors, VIMP enlarged the search field by using a safety search border of 
$1^\prime$ around the $(\alpha,\delta)$ positions listed in the NVO image archive. Due to this 
pointing incertitude, some VIs could actually fall outside the candidate fields. \\

Most false candidates were generated by the very optimistic limiting magnitude 
$V=26$ set to accommodate best sky conditions, predicted apparent magnitudes due to possible 
reported magnitudes and orbital incertitude, and possible larger light curve variations. Our
faintest detection was 2015~HO182 ($V=23.4$), which explains the large amount of VI candidates 
(212), which is four times higher than the actual number of encounters (54). \\

Many candidate encounters happened while rapid apparent motion and/or longer 
exposure times generated shorter or longer trails, which make the object invisible due
to the trail loss effect. About $30\%$ of VIs from the 54 detected objects presented trails,
and we can assume similar statistics for the invisible trails. \\

In some cases, the exposure times are quite shallow for fainter objects to
show up, which made some candidates invisible. \\

Some filters (especially $u$) are less sensitive to asteroid spectra, 
which make them invisible, even though they could be predicted to be brighter. \\

Some detections of star-like sources in singular images could be risky 
to report due to possible confusion with faint stars or background galaxy nuclei, and 
actually MPC does not recommend the reporting of one-night positions. Although we took all 
necessary steps to avoid such risks, in a few cases we needed to drop such candidate 
detections. 

\subsection{Orbital improvement for 54 VIs}

We reported a total of 183 positions for 54 objects (average 3.4 positions per object). 
Most p/recovered VIs have an actual Apollo classifications (43 or 80\%), while 6 are Amor 
and 5 are Atens. Our project produced data published in 52 MPC supplement publications. \\

Table~2 includes all 54 objects p/recovered by the VIMP DECam project. \\

Both star-like and star-like elongated apparitions (about half cases) were measured using 
the photocenter method, while the trail apparitions were measured calculating the middle point 
between the two ends or/and reporting both ends. \\

\subsubsection{Astrometry}

Accurate astrometry is essential for the orbital improvement and the eventual removal of VIs, 
and  we took some appropriate steps to ensure quality control for the correct object 
identification and position measurements. The archival DECam CCD images were found to be very flat, thus the initial DECam astrometry and the 
subsequent World Coordinates System (WCS) solution were found to be extremely accurate in most cases (except for very few fields 
located close to the equator that had wrongly indexed $\delta$ in the NVO archive). The initial DECam WCS 
is solved by the pipeline based on the Gaia-DR1 reference catalogue, using a sophisticated procedure 
that provides astrometric solutions within $0.03^{\prime\prime}$, dominated only by stochastic 
atmospheric distortion \citep{ber17}. \\

To identify targets and measure the astrometry, we used the Windows-based 
Astrometrica\footnote{www.astrometrica.at} software. This is the tool widely used by many  
amateur astronomers, including the two involved in the actual measurements of this project (LC and DC),
and many other volunteers involved in other similar EURONEAR projects. \\

In just a few cases when Astrometrica failed to converge (due to  $\delta$ image centres
close to equator that had been wrongly indexed by the NVO archive) or for double checking the increasing RMS of orbital fits, 
we used the online version of Astrometry.net to solve exactly the CCD centres expected as input 
in Astrometrica. We used Gaia-DR1 as the reference catalogue stars for all searched fields, typically identifying 
around 100 catalogue stars in each CCD with 
Astrometrica, which provided RMS plate fits using polinomial 
cubic fits within $0.1^{\prime\prime}$ in both $\alpha$ and $\delta$ in all cases. \\

Based on the improved orbital fit following our MPC VIMP reporting, in Fig.~8 we plot the 
current NEODyS ``observed minus calculated'' (O-C) residuals for all our measurements. 
Most O-Cs are confined to $0.3^{\prime\prime}$, with only 12 points above $0.5^{\prime\prime}$ in 
$\alpha$ or $\delta$, and the standard deviations are $0.22^{\prime\prime}$ in $\alpha$ and 
$0.17^{\prime\prime}$ in $\delta$. 

\begin{figure}
\centering
\includegraphics[angle=0,width=7.5cm]{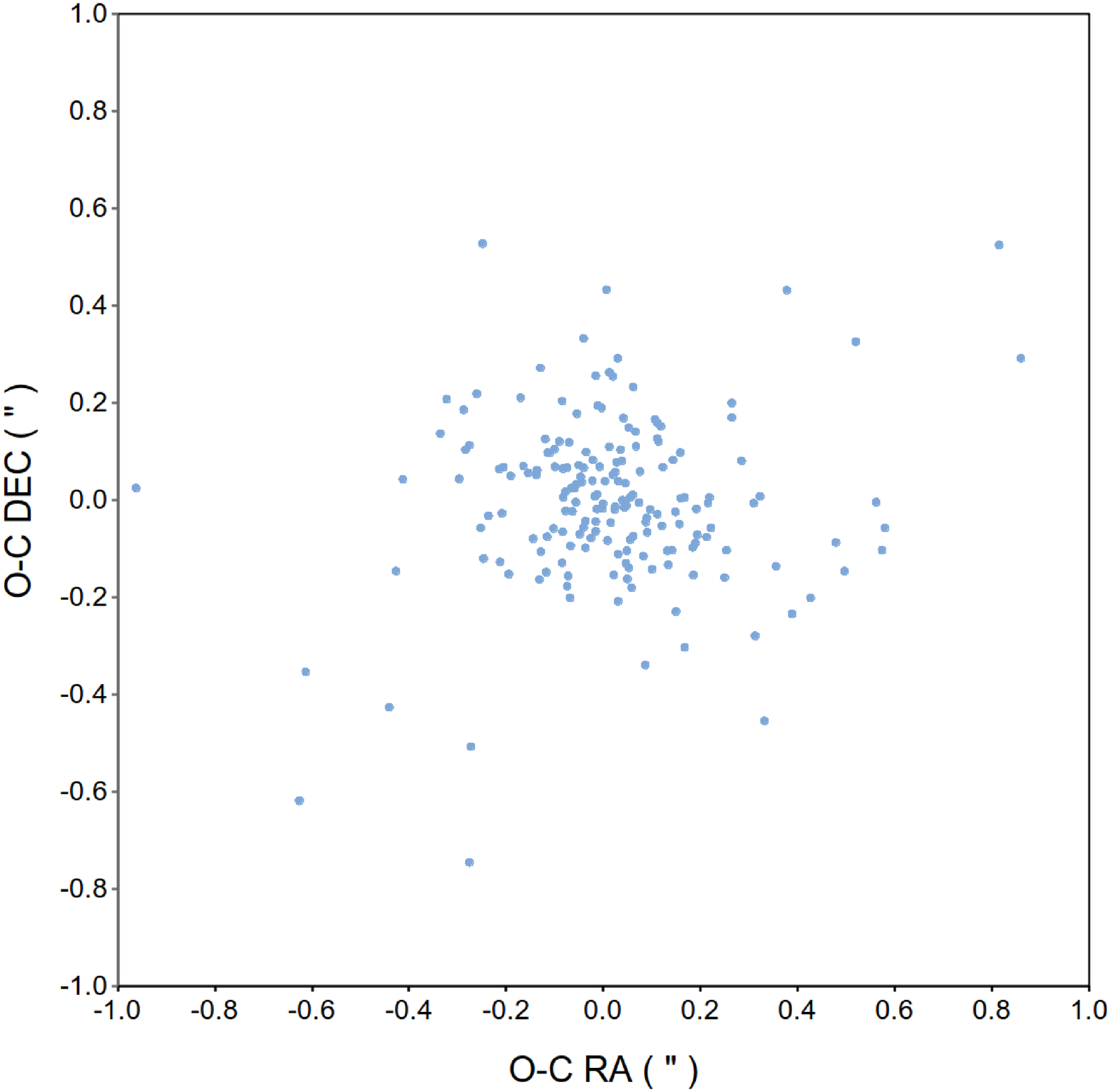}
\begin{center}
\caption{Observed minus calculated (O-C) residuals counting 183 measurements around the improved orbits 
         of 54 VIs p/recovered in the Blanco/DECam archive. } 
\label{fig8}
\end{center}
\end{figure}

\subsubsection{Orbital uncertainties}

The MPC orbital uncertainty factor $U$ (listed as a condition code by the JPL database) was reduced for 32 
VIs (59\% of the whole sample), counting seven very poorly observed objects (arcs of 1-2 days) with unknown 
$U$ before VIMP (marked by $-/$ in Table~2). 
The orbital uncertainty remained unchanged for 13 VIs (24\% of the entire sample).
Curiously, the orbital uncertainty increased for 7 VIs (13\%), namely: 2014~HN197 (5/9), 2014~JT79 
(5/7), 2014~JU79 (5/8), 2015~HV182 (7/8), 2016~WM1 (2/6), 2017~UK3 (6/7), and 2018~FY2 (8/9). \\

Seven of these increased orbital residuals could be due to apparitions falling close to the margins 
of the DECam field or close to CCD margins (possibly affected by larger field distortion), while two 
other cases could be due to the imperfect end of trails measurements of faint objects. Five of these seven 
objects had very short arcs before VIMP (1-2 days), which could affect the assessment of the orbital 
uncertainties. 

\subsubsection{Orbital amelioration}

We used Find\_Orb software\footnote{https://www.projectpluto.com/find\_orb.htm} (version 8 Oct 2019
installed under Linux Ubuntu 18) to check standard deviations of the orbital fits before and after VIMP
(column RMS in Table~2), removing all observations with larger $O-C>1^{\prime\prime}$ 
in either $\alpha$ or $\delta $ for each object, activating the Earth, Moon, other planetary, and major asteroids perturbers, 
then taking the ``full step'' least squares orbital fit method (which converges in very few steps). Nevertheless, 
Find\_Orb (including some older versions) could not fit elliptical orbits for the 
following objects: 2015~XP, 2016~CH30, 2017~EA, 2017~VC14, 2018~SD2, and 2019~JH7, for which we include 
RMS values from the MPC database. \\

We include the resulting orbital standard deviations in Table~2 (RMS column) before and after VIMP. 
For 16 VIs the RMS remains constant (30\%), for 24 objects the orbital RMS grows (44\%), and for 
14 VIs (26\%) it decreases. 
Typically RMS increases by $0.01-0.02^{\prime\prime}$. 
More substantial increases correspond to very poor observed objects (initial arc of 1-2 days), 
such as 2014~HN198, 2015~HE183, 2015~HW182, or 2018~SD2. Longer trailed objects do not 
contribute to the growth, proving the technique to average the two ends of the trails and our 
good visual approximation of the trail ends. 
The typical RMS decrease is $0.01-0.02^{\prime\prime}$, but a few VIs showed more substantial 
drops in RMS, namely 2016~JT38 (PHA with a very short, one-day arc prolonged to ten days) and 
2019~JH7 (with RMS values taken from MPC). \\

\subsubsection{Weird case of 2014~HG196}

Only the data set of one object (2014~HG196) from one night could not be fitted well in Find\_Orb 
and it was not reported to MPC after recovery. Our initial four-night recovery data sets (included 
in Table~2) fitted the older orbit very well (orbital mean residual $RMS=0.10^{\prime\prime}$ 
and extremely small individual residuals, mostly $O-C<0.10^{\prime\prime}$) in both $\alpha$ 
and $\delta$, enlarging the arc from one to eight nights. Later, we ran VIMP2 and recovered the object ($V=23.0$) during a fifth night (20140504 - 
not included in this paper) measuring three positions visible in three images that match the expected 
motion and are similar in appearance to previous reported VIMP nights, and show a clear 
detection based on track and stack, whose full width at half maximum matches the nearby star detections. During 
that night, the VI positional uncertainty ellipse is tiny ($a\sim5^{\prime\prime}$), thus 
the risk of confusion is extremely small. While adding this fifth night to increase the arc from 8 to 11 days, we noticed that 
Find\_Orb curiously jumped the orbit mean residual to $RMS=0.25^{\prime\prime}$, 
while all new measurements were systematically outside the previous arc (by 
$O-C_{\delta}=-0.86^\prime\prime$ and $O-C_{\alpha}=+1.52^\prime\prime$). 
We can not explain this result, despite trying alternative approaches (independent 
astrometric reduction and measurements, using different reference catalogues, adding solar 
radiation pressure, testing older versions of Find\_Orb, testing another software 
NEOPROP\footnote{http://neo.ssa.esa.int/neo-propagator}, checking for other known asteroids, 
investigating for a possible satellite), so recently we reported the fifth night data set to MPC.

\subsubsection{MOID improvement}

Besides the elimination of impact orbits, the minimum orbital intersection distance (MOID) represents 
the most important orbital parameter needed to assess risks \citep{mil09}. In Table~2 we include MOIDs 
before and after VIMP. MOID remains constant for 37 objects (67\% of cases), although impact orbits were 
eliminated for 15 of them from at least one list. MOID changed for 18 VIs, decreasing for 10 objects, with 
a substantial shrink for 2014~JU79, 2015~HW182 (both removed VIs), PHA 2016~JT38 and 2016~JO38 (which need 
follow-up to be removed), and increasing for 8 objects (with significant growth for 2015~HV182 and 2014~HN197, 
which are both PHAs). \\

\subsubsection{Precovery and recovery statistics}

Figure~9 shows the histogram of 54 VIs p/recovered in this VIMP DECam project, as a function of the diameters
of the objects, calculated from the absolute magnitude ($H$) assuming a spherical body with a visual 
albedo of 0.154 (in accordance with the Palermo Scale). Most objects (51 or 95\%) are smaller than 
100~m, with the smallest 14 (26\%) below 10~m and the largest three being 2014~HN197 (about 200~m, 
degraded by VIMP from PHA to NEA status due to revised MOID), 2015~HV182, and 2016~JT38 (both about 
300~m and still classified as PHAs). We underlined the designations of these objects in Table~2. 

\begin{figure}
\centering
\includegraphics[angle=0,width=9cm]{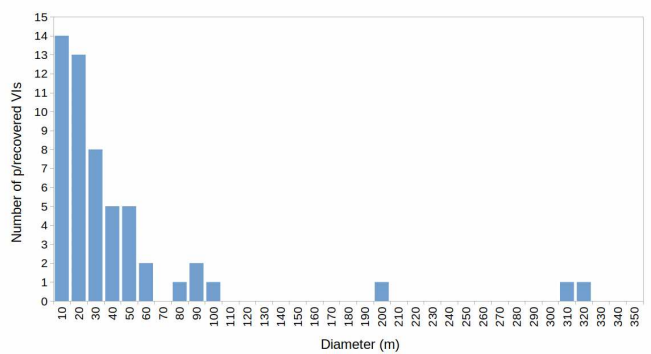}
\begin{center}
\caption{Histogram counting 54 VIs p/recovered in the Blanco/DECam archive function of their diameter, 
         assuming 0.154 albedos and spherical bodies. } 
\label{fig9}
\end{center}
\end{figure}

Figure~10 plots the histogram counting the first night apparitions for all VIs p/recovered by VIMP,
as a function of their predicted apparent magnitude $V$ based on actual MPC ephemerides. We chose to count
predicted $V$ in order to decrease errors with measured magnitudes due to different filters, unknown 
colours, and long trails whose magnitude could not be measured. The main bulk of detections fall 
between $21<V<23$, the faintest precovered objects being 2014~HT197 ($V=24.0$ during the first night 
in filter $VR$ for a 40~s exposure) and 2016~CH30 ($V=23.7$ detected in three images in the $r$ and $g$ bands 
using longer exposures), followed by another six objects in the next bin ($23.0<V<23.5$), thus apparently 
the Blanco/DECam limiting magnitude for asteroid data mining work is around $V\sim24.0$. 

\begin{figure}
\centering
\includegraphics[angle=0,width=9cm]{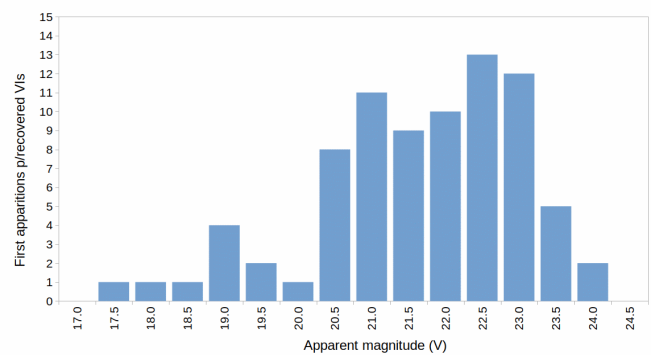}
\begin{center}
\caption{Histogram counting first night apparitions of the 54 VIs p/recovered in the Blanco/DECam archive
         as a function of their predicted apparent magnitude. } 
\label{fig10}
\end{center}
\end{figure}

Figure~11 plots two sets of histograms including all VIs p/recovered by VIMP in the DECam archive, 
based on two samples: the length of observed arcs before and after our VIMP work. 
The first sample (red hatched bars) counts the number of VIs based on the arcs observed before VIMP, 
while the solid blue bars count the number of VIs based on the arcs prolonged by VIMP (precovery and recovery). 
Most VIs had very short arcs previously: almost one third (16 objects or 30\%) with arcs shorter than one day, 
about half (30 objects or 56\%) with arcs shorter than three days, and three quarters (40 objects or 74\%) with 
arcs shorter than seven days. 
Most objects counted in the hatched bars in the left side of the histogram migrated randomly to the right
after p/recovery, increasing the heights of the blue bars. 
The shorted observed arc was 2017~EA (only four hours)\footnote{https://cneos.jpl.nasa.gov/news/news194.html}
, which was prolonged to 33 hours and removed from both lists following VIMP DECam data mining. \\

\begin{figure}
\centering
\includegraphics[angle=0,width=9cm]{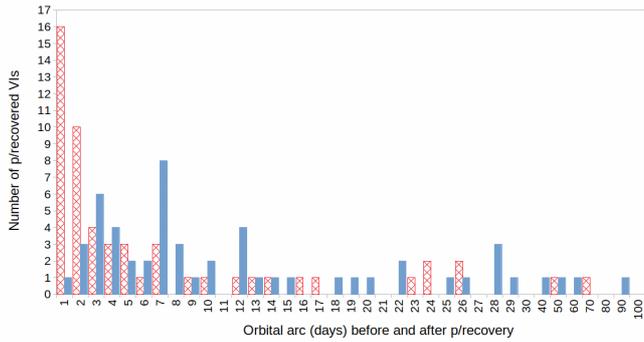}
\begin{center}
\caption{Histogram counting 54 VIs p/recovered in the Blanco/DECam archive as a function of their orbital arc, 
         before (hashed red) and after (solid blue) this work. 
         Please note the larger bin above X=30, in order to fit the longer observed objects. } 
\label{fig11}
\end{center}
\end{figure}

\subsection{Elimination of impact orbits}

Thanks to our VIMP DECam project, impact orbits for 27 VIs were eliminated from at least one risk list
(half of the entire sample). Fourteen of these objects were eliminated from both the CLOMON and Sentry lists 
(26\% of all 54 encountered objects). In Table~2 we mark in italics the VIs removed from 
one list (13 objects) and in bold the VIs removed from both the Sentry and NEODyS lists (14 objects). \\

Figure~12 counts all 54 VIs based on their initial arc (before our p/recovery) split into two samples: 
eliminated VIs (red hashed histogram) and remaining VIs (solid blue). Thirteen objects with very short initial 
arcs of less than two days were eliminated (24\% of a total of 54 VIs), and 19 objects with initial arcs of less than 
five days were eliminated (35\% of the entire sample). Compared with the total number of eliminated objects 
(27 VIs), these numbers correspond to about half (48\% corresponding to a two-day arc limit) and 
the majority of all eliminated objects (70\%). The best observed objects were 2014~EU (23 days), 2018~PZ21
and 2016~GS134 (24 days), 2018~FA4 and 2018~RY1 (26 days), 2018~TB (48 days), and 2018~EZ2 (66 days). \\

\begin{figure}
\centering
\includegraphics[angle=0,width=9cm]{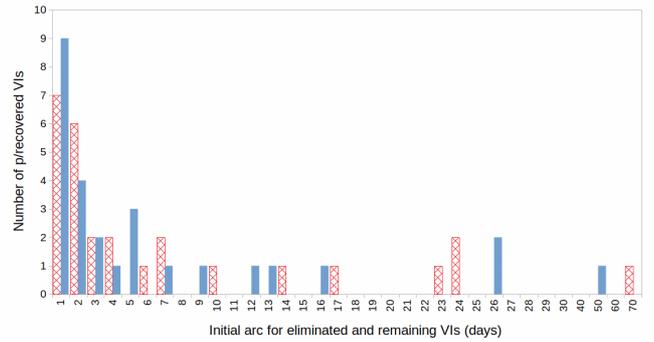}
\begin{center}
\caption{Histogram counting 54 VIs p/recovered in the Blanco/DECam archive as a function of their orbital arc, 
         grouped by eliminated (hashed red) and remaining VIs (solid blue). 
         Please note the larger bin above X=30, in order to fit the longer observed objects. } 
\label{fig12}
\end{center}
\end{figure}

Figure~13 plots the histogram for all VIs as a function of their arc extension, defined as the time interval 
between the initial and the final observed arc. 
The average extension of orbital arcs for the entire set is 5.5 days.
The majority of eliminated VIs (red hatched histogram), and remaining VIs (solid blue colour) had arc 
extensions of less than seven days, namely 21 objects (39\% of the entire sample or 78\% of all eliminated VIs) 
and 19 objects, respectively (35\% of the entire sample or 70\% of all remaining VIs). 
The longest arc extensions are 2018~EZ2 (by 19 days), 2016~JP38 (21 days), 2018~RY1 (23 days), and 2015~XP (25 days). \\

\begin{figure}
\centering
\includegraphics[angle=0,width=9cm]{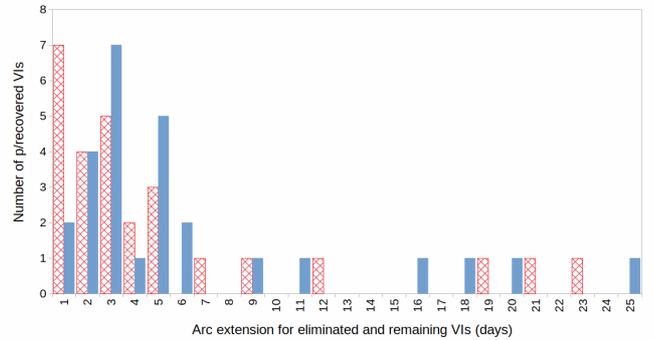}
\begin{center}
\caption{Histogram counting 54 VIs p/recovered in the Blanco/DECam archive function of their extension arc, 
         grouped by eliminated (hashed red) and remaining VIs (solid blue). } 
\label{fig13}
\end{center}
\end{figure}

\subsection{Removal of 27 VIs}

In Table~2 we mark in bold the VIs whose former impact orbits were eliminated by VIMP from both 
the CLOMON and Sentry lists, and in italics VIs with impact orbits eliminated from only one list. In 
columns $IOS$ and $IOC$ we list the number of impact orbits listed in the Sentry and CLOMON 
before and after VIMP (separated by $/$ symbol). In total, 27 VIs were removed from at least one list (13 from both lists and 14 from one list), 
and about 4000 impact orbits were eliminated in total from both lists (at least 690 from CLOMON 
and 3278 from Sentry). The number of impact orbits decreased or remained unchanged for 46 objects 
(85\% from all) in both lists after VIMP. For eight objects the number of impact orbits of at 
least one list increased after VIMP (2014~HN197, 2016~CH30, 2016~SA2, 2017~FB102, 2017~UK3, 
2018~BX5, 2018~EE9, and 2018~PZ21), so they need follow-up. \\

In four cases (2014~EU, 2014~HJ198, 2015~HE183, and 2015~XP) we failed to save the number of impact 
orbits before VIMP.
In another two cases (2017~SE33 and 2018~RY1), the Pan-STARRS F51 team that is mining the data from Pan-STARRS images 
precovered targets at the same time as our VIMP work (but they reported results first), and we did 
not save the impact orbits before VIMP. 
In another case (2019~MF3) we ran VIMP just two weeks after discovery but did not check the number
of impact orbits, and the object was removed thanks to new follow-up observations a few days later,  
so again we miss the impact orbits before VIMP. 
In all these six cases we mark the unknown number of impact orbits before VIMP by $?/$ in columns 
$IOS$ and $IOC$. \\

The three largest VIs ($>100$~m, thus PHAs) that could not be removed are 2014~HN197, 2015~HV182,
and 2016~JT38, so additional data mining or future recovery is needed for them. Nevertheless, 
2014~HN197 was degraded from the PHA to the NEA list (due to an increase in MOID). \\

\subsection{Fast rotator VIs}

The rotational properties of small NEAs are important for providing clues about the formation 
and evolution of asteroids. However, the study of bodies smaller than 100~m requires very large 
telescopes and extended periods to accommodate the multiple exposures needed to determine their 
light curves using the classic method. An alternative, quick way could be the inspection of 
trailing objects that fall serendipitously into longer exposed, data-mined images, which can reveal 
very fast rotators. For example, during the discovery of 2014~NL52 (second EURONEAR NEA 
discovery) we also found its rapid rotation period of 4.453$\pm$0.003 min \citep{vad15}.\\

Some surprising photometric results were obtained during our VIMP project upon careful visual 
inspection of the opportunities generated by longer exposures or/and fast-moving VIs. During this 
project we found 16 targets presenting trails longer than 50 pixels. The image examination of at 
least three of these VIs strongly suggests very fast rotation. These are 2017~SQ2 (found on two 
exposures), 2018~TB, and 2018~DB. Further on, another six VIs are possible candidates for rapid 
spinning, found by analysing the profile of the trails.  Consequently, we have envisioned a 
new ``light curve trailing method'' to attempt period fitting for such cases. As 
photometric analysis was outside the aim of our VIMP project, we only show the preliminary 
results for 2018~DB here, and we will present a complete analysis in a future paper . \\

In Table~2 we include the observing circumstances of 2018~DB, a VI of about 10~m size 
(assuming an albedo of 0.154). This object was recovered in only one DECam image exposed for 200~s 
one day following its last observation, moving very fast ($\mu=133.81^{\prime\prime}$/sec) 
and leaving a very long trail (1697 pixels based on the NEODyS ephemerides), which ends in the 
gap between CCDs 7 and 8.  Moreover, the start of the trail falling in CCD 7 was difficult 
to assess, due to some evident periodic drop in faintness along the trail. Because of these 
reasons, we refrained from reporting any position to MPC. Nevertheless, the inspection of this 
frame clearly revealed a few repeating drops in brightness along the trail (Fig.~14) 
alternating with seven maxima whose measured positions and known proper motion could result 
in a first approximation of the rotation period. \\

\begin{figure}
\centering
\includegraphics[angle=0,width=9cm]{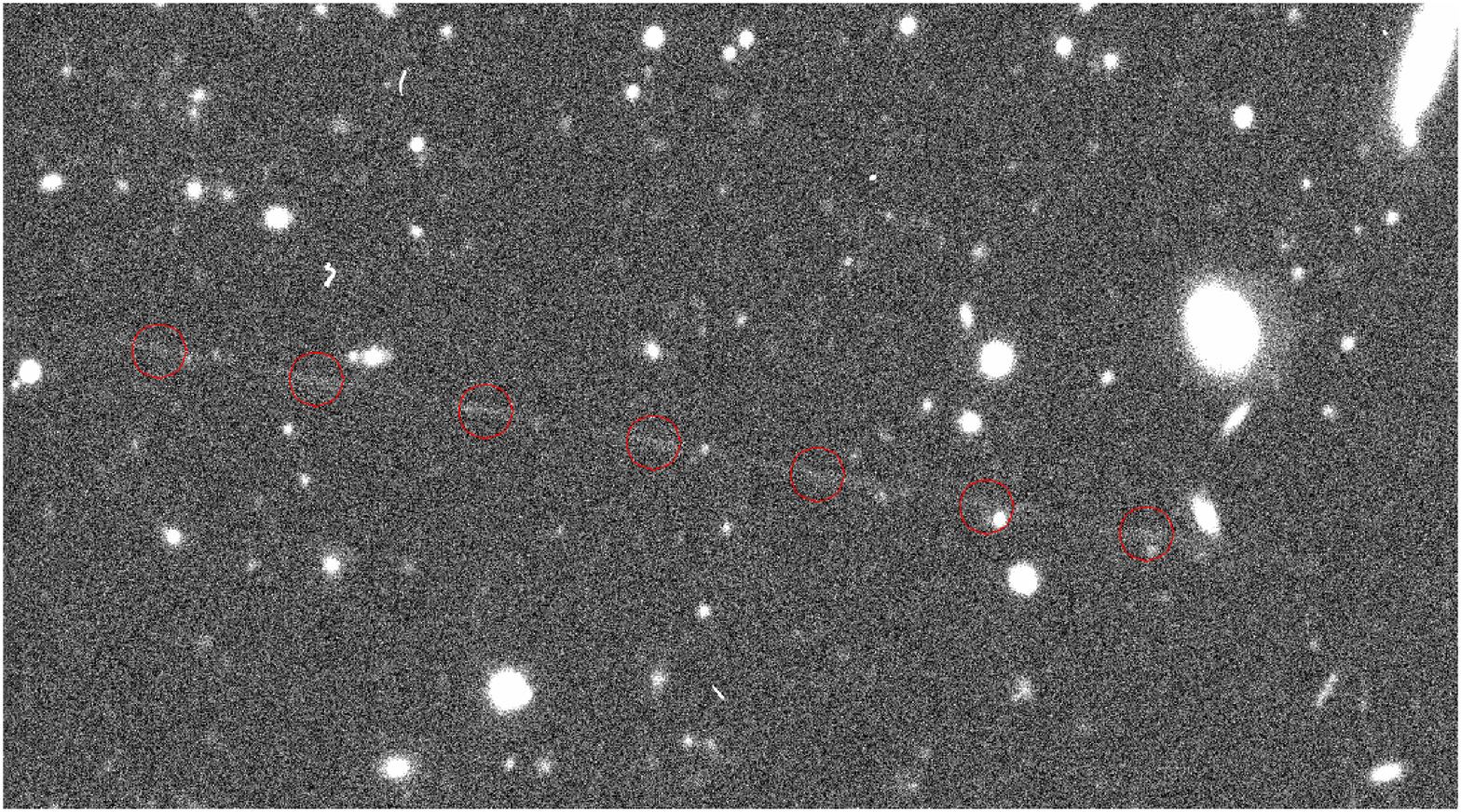}
\begin{center}
\caption{Trailing apparition of VI 2018~DB in one DECam image clearly showing periodic flux 
         variation along the very long trail. We mark with circles the seven maxima whose positions 
         and known proper motion could demonstrate its very rapid rotation period. }
\label{fig14}
\end{center}
\end{figure}

To assess the period accurately, we measured the flux along the trail using a 
{\it GNU Octave}\footnote{https://www.gnu.org/software/octave/doc/v4.4.1/} \citep{eat18} script 
developed in-house. The program steps the crossing line pixel by pixel, sampling the maximum flux
along the trail. For each of these sampled points along the trail, the corresponding flux can be
measured as the sum of fluxes in pixels in a rectangular aperture perpendicular to the trail, 
one pixel in width and seven to ten pixels in length (adjustable based on the seeing). The 
value of the background level sampled outside this aperture is subtracted, considering the median 
value of a region of 30 pixels outside of the given aperture (on both sides). The final step for 
obtaining the light curve consists in the calibration of the time axis, taking into account the pixel 
scale and the target proper motion. \\

Following this method, here we present the results for 2018~DB. The detected trail could be 
visually detected across a trail 1000 pixels long, with margins that were difficult to assess due 
to the limited S/N and the rapid flux variation. In order to improve the detection, 
we applied a binning of five points along the light curve. Then, we transformed the flux to magnitudes 
using an arbitrary zero point. Finally, by using the 
MPO Canopus\footnote{http://www.minorplanetobserver.com/MPOSoftware/MPOCanopus.htm} software, 
we succeeded in fitting a very fast rotation period $P=0.0082\pm0002$~h (29.5~s) for 2018~DB (Fig.~15), 
despite the very large error bars, dropping only several outlier points in which measured fluxes were 
below the sky. This bonus photometric result of our VIMP project makes 2018~DB the fastest known 
virtual impactor and the ninth fastest known asteroid, according to the MPML asteroid light curve 
database\footnote{http://www.minorplanet.info/PHP/lcdbsummaryquery.php}. 

\begin{figure}
\centering
\includegraphics[angle=0,width=9cm]{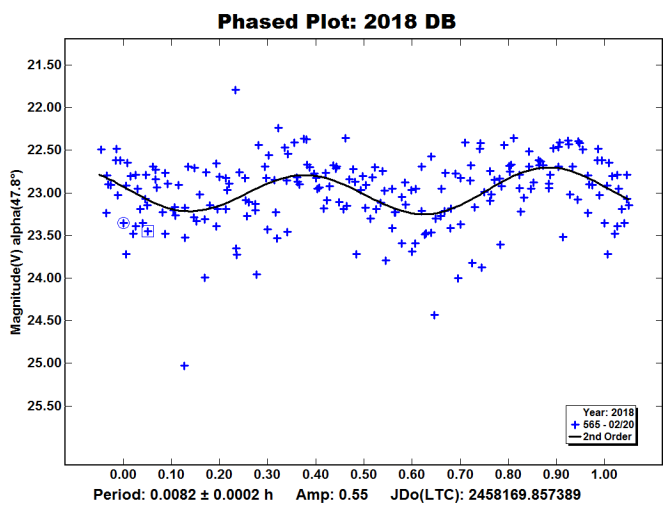}
\begin{center}
\caption{Trailing light curve method (to be presented in a future paper) applied 
         to the very long trail of the VI 2018~DB. We succeed in fitting $P=0.0082\pm0002$~h (29.5~s) 
         rotation period, despite the very large error bars not shown in the fit in order to avoid 
         over-plotting. }
\label{fig15}
\end{center}
\end{figure}

%______________________________________________________________

\section{Conclusions and future work}

Big data mining of image archives observed with large etendue facilities represents a great 
opportunity for the orbital amelioration of poorly known NEAs and VIs that serendipitously appear, 
and could be searched for, in these images, at no additional telescope cost. In light of this, we propose the 
Virtual Impactor search using Mega-Precovery (VIMP) 
project, which developed a Python software endowed with an effective (fast and accurate) 
algorithm for predicting apparitions of candidate pairs (asteroid, image) for a subsequent 
guided human search by data miners, which could involve citizen scientists. Our main results are as follows: \\

\begin{itemize}

\item[$\star$] We applied the VIMP software to mine the data of 451,914 Blanco/DECam images 
observed between 12 September 2012 and 11 July 2019, 
searching for about 920 VIs that appeared on the Sentry or CLOMON risk lists by 11 July 
2019. 

\item[$\star$] Imposing some quite relaxed search criteria (to avoid losses), VIMP
identified 212 VI candidates possibly appearing in 1286 candidate images leading to either 
precovery (before discovery) or recovery events (following the last reported observation). 

\item[$\star$] Fifty-four VIs were p/recovered in 183 DECam images (about 3.4 positions/object),
meaning a relatively low success rate (25\%) due to our relaxed search criteria. 

\item[$\star$] Considering a simple geometrical model, the VIMP algorithm 
searches for any possible intersection in space and time between the positional uncertainty 
(approximated by an ellipse) of a poorly known asteroid (such as a VI) and the bounding sky 
projection of any given archive image (approximated by a rectangle). 

\item[$\star$] The MPC orbital uncertainty factor was reduced for 32 VIs (59\%), remaining 
unchanged for an other 13 VIs (29\% of all encountered objects). 

\item[$\star$] Thanks to the very accurate Gaia-DR2 astrometry, the very good seeing of Cerro
Tololo, and the very good optical performance of DECam, most O-C residuals are confined within 
$0.3^{\prime\prime}$ , with standard deviations of $0.22^{\prime\prime}$ in $\alpha$ and 
$0.17^{\prime\prime}$ in $\delta$.

\item[$\star$] Using Find\_Orb software and the MPC observations database, we assessed 
the orbital standard deviations before and after VIMP. For 16 VIs the RMS remains constant 
(30\%), for 24 objects the orbital RMS grows (44\%), and for 14 VIs (26\%) it decreased a 
bit. 

\item[$\star$] The minimal orbital intersection distance remains constant for 37 objects (67\%), decreases for 
10 objects, with major drops for 2014~JU79, 2015~HW182 (both removed VIs) and the PHAs 2016~JT38 
and 2016~JO38, which need additional VIMP2 data mining using other archives or future 
observing), and increases for 8 objects (with significant growth for PHAs 2015~HV182 and 
2014~HN197). 

\item[$\star$] Most p/recovered objects (51 or 95\%) are smaller than 100~m, the smallest
below 10~m (14 objects or 26\%) and the largest being 2014~HN197 (about 200~m, degraded by 
VIMP from PHA to NEA status), 2015~HV182, and 2016~JT38 (PHAs of about 300~m diameter). 

\item[$\star$] The faintest detections were about $V\sim24.0$, and the main bulk falls 
between $21<V<23$ mag. 

\item[$\star$] Impact orbits for 27 VIs were eliminated from at least one risk list 
(half of the entire p/recovered sample), including 14 objects eliminated from both lists 
(26\% of all 54 encountered objects). Many eliminated VIs (70\%) had short initial arcs of
less than five days. 

\item[$\star$] Overall, about 4000 impact orbits were eliminated in total from both 
lists (at least 690 from CLOMON and 3278 from Sentry). The number of impact orbits
decreased or remained unchanged for 46 objects (85\%) in both lists after VIMP. 
For eight objects the number of impact orbits of at least one list increased (2014~HN197,
2016~CH30, 2016~SA2, 2017~FB102, 2017~UK3, 2018~BX5, 2018~EE9, and 2018~PZ21, which need 
follow-up). 

\item[$\star$] The three largest VIs could not be removed (2014~HN197, 2015~HV182, 
and 2016~JT38, all needing follow-up), nevertheless the first was degraded from the PHA 
to the NEA list (due to an increase in the MOID).

\item[$\star$] Many p/recovered objects were actually discovered by Blanco/DECam and 
could be linked to three runs (30 nights in total) carried out over three years for the NEO Survey 
program (PI: Lori Allen), counting 21 objects in total (39\% of all Table~2) from which 
we removed 16 VIs (including 9 from both lists).

\item[$\star$] Four other archives (big etendue facilities and number of images) 
were identified and queried by VIMP, producing encouraging results for other future searches, 
namely: CFHT/MegaCam (58 candidate VIs and 356 candidate images), Pan-STARRS1 (22 VIs and 
51 images), Subaru/HSC (28 VIs and 248 images), and VISTA/VIRCAM (17 VIs in 533 candidate 
images). 

\item[$\star$] Some surprising photometric discoveries have emerged from the project, 
based on the close inspection of longer trailed VIs after measuring their fluxes along 
the trails. These include the very fast rotator 2018~DB for which we derived rotation 
period $P=30$s, which makes this 10~m object the fastest known VI and the ninth most 
rapid asteroid known to date. In a future paper we will publish the method to reduce 
such trailed occurrences, and apply it to other similar DECam apparitions to derive 
probable fast rotators or candidates. 

\end{itemize} 

In the near future we plan a few improvements of the algorithm in order to reduce the number 
of false candidates and ease the manual search process, aiming to involve other students 
and amateur astronomers in the first two: 

\begin{itemize}

\item[$\star$] Search all VIs for possible apparitions inside the actual observed arc. 

\item[$\star$] Search all VIs in the other four instrument archives from Table~1. 

\item[$\star$] Search larger objects not yet removed in this project in all other archives
available in the EURONEAR Mega-Archive database (currently 112 instruments). 

\item[$\star$] Take into account the Moon (phase and separation) based on date, time, and the 
observing location, and decrease the limiting magnitude for affected candidate images. 

\item[$\star$] Use proper motion (queried in the ephemerides) to assess the drop in visibility 
for fainter objects based on the trailing loss effect. 

\item[$\star$] Take into account exposure time and filters to evaluate actual limiting 
magnitudes. 
 
\item[$\star$] For mosaic cameras, decompose the approximate rectangular bounding in individual 
CCDs with geometry defined in some instrument configuration files already used in other 
EURONEAR projects like NEARBY \citep{vad19b}. 

\item[$\star$] The great majority of VIs are small and faint, thus they could be p/recovered 
only while passing close to Earth. Based on this fact, better known objects can be searched 
only around such close approaches when they get brighter, to decrease the VIMP execution time. 

\end{itemize}

%______________________________________________________________

\begin{acknowledgements}

This project used data obtained with the Dark Energy Camera (DECam), which was 
constructed by the Dark Energy Survey (DES) collaboration. 
We made use of the U.S. National Virtual Observatory (NVO) for extracting the entire
DECam image archive and download the image candidates. 
Throughout the whole project we used the Near Earth Objects - Dynamic Site 
(NEODyS-2) whose ephemerides and positional uncertainties are essential to guide the 
search of asteroids serendipitously falling in archival images. 
To measure astrometry we used the Astrometrica software written by Herbert Raab. 
We used Aladin desktop and a small bash script to cut candidate CCDs from the mosaic image. 
In a few cases we used the Astrometry.net online to check the accurate centres of
a few CCD images. 
In very few cases we used IRAF to correct some CCD bias levels possibly caused sometimes 
by different amplifiers. 
The Find\_Orb orbital fitting software of Bill Gray was essential for confirming 
correct pairing of the searched objects, the last quality control step before reporting 
our data to MPC. 
We used the SAO DS9 software for overlaying the uncertainty region files generated by 
VIMP, and in a few cases we employed IRAF to equalize the two-level bias levels present 
in some CCDs. 
The Romanian amateur astronomers Daniel Bertesteanu (Bucharest) and Iulia Dima (Timisoara) 
got involved for a while in this project for training and few trial measurements. 
Thanks are due to the anonymous referee whose comments helped us improved the paper. 

\end{acknowledgements}

%______________________________________________________________

% \clearpage
% Figures

%__________________________________________

% TABLES

\clearpage
\onecolumn
\begin{table}[t]
\begin{center}
\caption{Large etendue instruments queried for VIs within VIMP project. } 
\label{table1}
\begin{tiny}
\begin{tabular}{lrrrrrrrrrr}
\noalign{\smallskip}\hline\hline\noalign{\smallskip}
Telescope   &  Camera  & Nr CCDs &  $V$ &  FOV &  $A\Omega$ & Nr imgs & Obs dates & Cand VIs & Cand imgs \\
\noalign{\smallskip}\hline\hline\noalign{\smallskip}
Blanco      &    DECam &    62   &  26  &  2.9 &   33 &   451,914 &  20120912-20190711 &   212   & 1286  \\
CFHT        &  MegaCam &    36   &  25  &  1.0 &    8 &   248,419 &  20030222-20191029 &    58   &  356  \\
Pan-STARRS1 &  PS1/DR2 &    60   &  23  &    7 &   12 &   374,232 &  20090603-20150225 &    22   &   51  \\
Subaru      &      HSC &   104   &  26  &  1.9 &   95 &    57,712 &  20140326-20180322 &    28   &  248  \\
VISTA       &   VIRCAM &    16   &  24  &  0.6 &    6 & 1,745,316 &  20091016-20200317 &    17   &  533  \\
\noalign{\smallskip}\hline\hline\noalign{\smallskip}
\end{tabular}
\end{tiny}
\end{center}
\end{table}

\clearpage
\onecolumn
%\begin{sidewaystable}
\begin{tiny}
\begin{landscape}
\label{table2}
\begin{center}
%\begin{tabular}{lrrrrrrrrrrrrrrrr}
\begin{longtable}{lrrrrrrrrrrrrrrrrrrrr}
\caption{
Virtual impactors p/recovered in the Blanco/DECam. 
The columns list designations, orbital class (APollo, AMor, or ATen), the discovery site
(PC code), diameter (in metres) assuming an albedo of 0.154, VIMP encounter event (Precovery or 
Recovery), image date (YYYYMMDD format), predicted apparent magnitude 
($V$ based on the actual MPC ephemeride), number of reported positions ($n$), image exposure
time (Texp in seconds), image filter band, asteroid proper motion ($\mu$ in $^{\prime\prime}$/min)
trail length (in DECam pixels), observed arc (in days, before and after our VIMP search), 
standard deviation of the orbital fit before and after VIMP (RMS in $0.01\arcsec$), Earth minimal 
orbital intersection distance before and after VIMP (MOID, in $0.0001$ a.u.), MPC orbital 
uncertainty before and after VIMP, number of impact orbits before and after VIMP ($IOS$ as listed 
by Sentry, and $IOC$ by CLOMON), the reducer initials (LC and DC), some comments, and MPC 
publication (MPS). 
The Comments column includes the following acronyms regarding the apparition of searched VIs
(with arbitrary limits): sl = star-like, sle = star-like elongated (bellow 15 pixels), 
st = small trail (15-50 pix), lt = long trail (50-200 pix), vlt = very long trail (200-1000 pix), 
elt = extremely long trail (above 1000 pix), 
met = reporting middle ends of trail, emet = reporting ends and middle ends of trail, 
ts = measured using track and stack image co-addition, 
F51 or T14 = p/recovered by other teams in Pan-STARRS1 or CFHT images during our VIMP project,
and nr = not reported. 
} \\
\hline\hline\noalign{\smallskip}
VI          & Class & Disc & Diam & Event & Date &  $V$ & $n$ & Texp& Band & $\mu$ & Trail & Arc &   RMS   &    MOID  & $U$ & $IOS$ & $IOC$ & Red & Comments & MPS   \\
\noalign{\smallskip}\hline\hline\noalign{\smallskip}
\endfirsthead
\caption{Continued.}\\
\hline\hline\noalign{\smallskip}
VI          & Class & Disc & Diam & Event & Date &  $V$ & $n$ & Texp& Band & $\mu$ & Trail & Arc &   RMS   &    MOID  & $U$ & $IOS$ & $IOC$ & Red & Comments & MPS   \\
\noalign{\smallskip}\hline\hline\noalign{\smallskip}
\endhead
\endfoot
\noalign{\smallskip} 
{\it 2013 TH6}   &   AP  &  G96 &  23 & P & 20130930 & 18.9 &   1 &  90 &    r &  32.2 &   200 &  7/13 & 26/26 &    79/79 &  7/7 &   --- &  1/0  &LC,DC& vlt,oet &1001151 \\
\noalign{\smallskip} 
2014 EU          &   AP  &  703 &  10 & P & 20140305 & 19.0 &   2 & 180 &    g &  12.7 &   144 & 23/24 & 24/23 &    26/26 &  6/4 &   9/5 &  ?/7  &  LC &  lt,met & 915088 \\
\noalign{\smallskip} 
{\it 2014 HD199} &   AP  &  W84 &  81 & R & 20140502 & 20.5 &   4 &  40 &   VR &   1.7 &     4 &   2/4 &  9/14 & 915/1032 &  9/8 &   2/0 &  ---  &  LC & sl      & 942498 \\
\noalign{\smallskip} 
{\it 2014 HG196} &   AP  &  W84 &  30 & R & 20140425 & 22.5 &   5 &  40 &   VR &   0.3 &     1 &   1/8 & 26/26 &    79/79 &  5/5 & 329/0 &  ---  &  LC & sl      &1034029 \\
           ...   &  ...  &  ... & ... & R & 20140426 & 22.3 &   3 &  40 &   VR &   0.4 &     1 &   ... &  ...  &      ... &  ... &   ... &  ...  &  LC & sl      &1034029 \\
           ...   &  ...  &  ... & ... & R & 20140430 & 21.3 &   3 & 300 &   VR &   1.2 &    23 &   ... &  ...  &      ... &  ... &   ... &  ...  &  LC & st      &1149506 \\
           ...   &  ...  &  ... & ... & R & 20140501 & 21.0 &   3 & 300 &   VR &   1.7 &    32 &   ... &  ...  &      ... &  ... &   ... &  ...  &  LC & st      &1149506 \\
\noalign{\smallskip} 
2014 HJ198       &   AP  &  W84 &  17 & P & 20140424 & 20.1 &   4 &  40 &   VR &   8.6 &    22 &   1/3 & 12/13 &    81/79 &  8/5 & 124/1 &  ?/4  &  LC & st,met  & 919176 \\
           ...   &  ...  &  ... & ... & P & 20140425 & 20.6 &   5 &  40 &   VR &   5.1 &    13 &   ... &   ... &      ... &  ... &   ... &  ...  &  LC & st,met  & 919176 \\
\noalign{\smallskip} 
{\it 2014 HK197} &   AP  &  W84 &  16 & P & 20140423 & 22.6 &   2 &  40 &   VR &   2.0 &     5 &   1/2 & 12/12 &  258/258 &  9/7 &  56/0 & 17/3  &  LC & sl      & 974491 \\
\noalign{\smallskip} 
{\bf 2014 HN198} &   AP  &  W84 &  37 & R & 20140502 & 22.2 &   3 &  40 &   VR &   1.3 &     3 &   1/6 &  7/14 &  205/193 &  4/7 &  97/0 &148/0  &  LC & sl      & 969838 \\
           ...   &  ...  &  ... & ... & R & 20140503 & 22.4 &   3 &  40 &   VR &   1.2 &     3 &   ... &   ... &      ... &  ... &   ... &  ...  &  LC & sl      & 969838 \\
\noalign{\smallskip} 
{\bf 2014 HN199} &   AP  &  W84 &  15 & R & 20140430 & 21.5 &   3 &  40 &   VR &  14.2 &    36 &   4/7 & 10/22 &  191/210 &  6/4 &  16/0 &  9/0  &  LC & st,met  & 940850 \\
\noalign{\smallskip} 
{\bf 2014 HS197} &   AP  &  W84 &   5 & P & 20140423 & 21.7 &   3 &  40 &   VR &  15.6 &    39 &   1/2 & 40/39 &    62/62 &  6/4 &   7/0 & 11/0  &  LC & st,F51  & 928512 \\
\noalign{\smallskip} 
\underline{2014 HN197}&AP&  W84 & 194 & P & 20140423 & 22.7 &   4 &  40 &   VR &   1.0 &     3 &   1/2 &   6/6 &   80/689 &  5/9 &  2/57 & 59/51 &  LC & sl      & 974154 \\ % U and IOS growed!
\noalign{\smallskip} 
{\bf 2014 HT197} &   AP  &  W84 &   6 & P & 20140423 & 24.0 &   2 &  40 &   VR &   1.9 &     5 &   2/3 & 12/11 &    36/36 &  8/7 &   4/0 &  2/0  &  LC & sl      & 986649 \\
           ...   &  ...  &  ... & ... & R & 20140430 & 22.0 &   2 & 300 &   VR &  13.7 &   259 &   ... &   ... &      ... &  ... &   ... &  ...  &  LC & vlt,met & 986649 \\
\noalign{\smallskip} 
{\bf 2014 JT79}  &   AP  &  W84 &  27 & P & 20140428 & 21.8 &   3 &  40 &   VR &   3.5 &     9 &   3/7 & 23/22 &    58/58 &  5/7 &1130/0 &  7/0  &  LC & sle,ts  & 994350 \\
\noalign{\smallskip} 
{\it 2014 JU79}  &   AP  &  W84 &  98 & P & 20140504 & 23.0 &   2 &  40 &   VR &   1.3 &     3 &   1/2 & 11/14 &   906/48 &  5/8 &  10/0 & 18/1  &  LC & sl,ts   & 986651 \\ % U growed! 
\noalign{\smallskip} 
2014 MO68        &   AP  &  F51 &  73 & P & 20140620 & 21.3 &   6 &  20 &   gr &   1.9 &     2 &  1/11 & 11/12 &    91/90 &  -/8 & 203/2 & 44/3  &  LC & sl      &1001200 \\
           ...   &  ...  &  ... & ... & P & 20140620 & 21.3 &   1 &  30 &    Y &   1.9 &     4 &   ... &   ... &      ... &  ... &   ... &  ...  &  LC & sl      &1001200 \\
\noalign{\smallskip} 
2014 UU56        &   AP  &  G96 &   7 & P & 20141021 & 19.0 &   2 &  90 &   gr &  13.5 &    77 &   4/7 & 22/22 &     4/4  &  6/6 &   --- &  4/2  &  LC & lt,et   & 942508 \\
\noalign{\smallskip} 
2015 HE183       &   AP  &  F51 &   8 & P & 20150421 & 19.2 &   4 &  40 &   VR &  48.9 &   123 &   2/5 & 16/32 &    42/42 &  9/6 &   3/1 &  ?/1  &  LC & lt,met  & 915198 \\ % Mega-Prec from Design can't find it! (but from orbit OK)
\noalign{\smallskip} 
2015 HO182       &   AP  &  W84 &  18 & R & 20150421 & 23.3 &   4 &  40 &   VR &   1.7 &     4 &   2/5 &  7/9  &  397/303 &  8/7 &  30/6 & 32/6  &  LC & sl      & 992726 \\
           ...   &  ...  &  ... & ... & R & 20150422 & 23.4 &   2 &  40 &   VR &   1.6 &     4 &   ... &  ...  &      ... &  ... &   ... &  ...  &  LC & sl,ts   & 992726 \\
\noalign{\smallskip} 
{\bf 2015 HW182} &   AM  &  W84 &  32 & P & 20150415 & 23.2 &   1 &  40 &    r &   1.2 &     3 &   1/4 &  7/10 & 1254/661 &  9/5 & 177/0 &  9/0  &  LC & sl      & 999847 \\
           ...   &  ...  &  ... & ... & P & 20150415 & 23.2 &   1 &  49 &    g &   1.2 &     4 &   ... &   ... &      ... &  ... &   ... &  ...  &  LC & sl      & 999847 \\
\noalign{\smallskip} 
{\bf 2015 HS182} &   AP  &  W84 &  43 & R & 20150422 & 22.2 &   5 &  40 &   VR &   1.9 &     5 &   3/4 &   8/9 &  241/237 &  9/8 & 113/0 & 38/0  &  LC & sl      & 952798 \\
\noalign{\smallskip} 
\underline{2015 HV182}&AM&  W84 & 310 & P & 20150415 & 21.5 &   4 &  40 &   gr &   1.2 &     3 &   1/7 &   6/8 &   57/257 &  7/8 & 189/8 &118/5  &  LC & sl      & 952798 \\
           ...   &  ...  &  ... & ... & R & 20150421 & 21.9 &   4 &  40 &   VR &   1.1 &     3 &   ... &   ... &      ... &  ... &   ... &  ...  &  LC & sl      & 952798 \\
\noalign{\smallskip} 
{\bf 2015 KA158} &   AP  &  W84 &  20 & P & 20150521 & 22.6 &   4 &  40 &   VR &   1.9 &     5 &   2/3 &  8/10 &  247/247 &  9/6 &  36/0 & 15/0  &  LC & sl      & 952802 \\
\noalign{\smallskip} 
2015 XP          &   AP  &  703 &  24 & P & 20151107 & 22.9 &   3 &  90 &  irg &   0.7 &     4 &  3/28 & 35/35 &      3/3 &  6/1 &   1/1 &  ?/1  &  LC & sl,pb   & 915217 \\
           ...   &  ...  &  ... & ... & P & 20151115 & 22.3 &   3 &  90 &  rig &   0.9 &     5 &   ... &   ... &     ...  &  ... &   ... &  ...  &  LC & sl,pb   & 915217 \\
\noalign{\smallskip} 
2016 CH30        &   AT  &  F51 &   9 & P & 20160117 & 23.7 &   1 &  52 &    r &   0.3 &     1 &  4/21 & 37/35 &    29/29 &  -/3 &   3/5 &  8/9  &LC,DC& sl      &1001388 \\
           ...   &  ...  &  ... & ... & P & 20160117 & 23.7 &   1 &  99 &    g &   0.3 &     2 &   ... &   ... &      ... &  ... &   ... &  ...  &LC,DC& sl      &1001388 \\
           ...   &  ...  &  ... & ... & P & 20160117 & 23.7 &   1 &  80 &    g &   0.3 &     2 &   ... &   ... &      ... &  ... &   ... &  ...  &LC,DC& sl      &1001388 \\
\noalign{\smallskip} 
2016 GS134       &   AP  &  F51 &  10 & R & 20160501 & 22.7 &   2 &  40 &   VR &   1.2 &     3 & 24/29 & 25/25 &     8/8  &  3/1 &   2/2 &  6/6  &  LC & sl,ts   &1012774 \\
\noalign{\smallskip} 
{\it 2016 JG38}  &   AP  &  W84 &  51 & R & 20160504 & 22.5 &   3 &  40 &   VR &   1.6 &     4 &   2/3 &  7/9  &  156/183 &  -/6 & 127/0 & 1/2   &  LC & sl      & 969875 \\ % ION growed!
\noalign{\smallskip} 
{\it 2016 JL38}  &   AP  &  W84 &  23 & P & 20160501 & 22.0 &   2 &  40 &   VR &   2.6 &     7 &   1/3 & 13/15 &  311/305 &  -/8 & 108/0 &  ---  &  LC & sl,F51  & 969875 \\
\noalign{\smallskip} 
\underline{2016 JT38}&AP &  W84 & 312 & P & 20160426 & 20.8 &   4 & 300 &   VR &   0.9 &    17 &  1/10 & 37/33 &   259/77 &  -/6 &  29/2 & 16/4  &  LC & st      & 940395 \\
           ...   &  ...  &  ... & ... & P & 20160427 & 20.7 &   2 & 300 &   VR &   1.0 &    19 &   ... &   ... &      ... &  ... &   ... &  ...  &  LC & st      & 940395 \\
           ...   &  ...  &  ... & ... & P & 20160501 & 20.2 &   3 &  40 &   VR &   1.1 &     3 &   ... &   ... &      ... &  ... &   ... &  ...  &  LC & sl      & 940395 \\
           ...   &  ...  &  ... & ... & P & 20160503 & 20.2 &   2 &  40 &   VR &   1.2 &     3 &   ... &   ... &      ... &  ... &   ... &  ...  &  LC & sl      & 940395 \\
\noalign{\smallskip} 
{\bf 2016 JP38}  &   AM  &  W84 &  51 & P & 20160501 & 22.1 &   4 &  40 &   VR &   0.3 &     1 &   1/3 &  6/7  & 2237/2520&  -/6 &  29/0 & 11/0  &  LC & sl,F51  & 928652 \\ % U NA ??
\noalign{\smallskip} 
2016 JO38        &   AP  &  W84 &  41 & P & 20160427 & 23.0 &   2 & 300 &   VR &   1.0 &    19 &   1/7 &  8/11 &   130/31 &  -/8 & 454/1 & 95/1  &  LC & sl,ts   &1014924 \\
\noalign{\smallskip} 
2016 SA2         &   AP  &  703 &   8 & R & 20161008 & 22.5 &   3 & 420 &   VR &   0.4 &    11 & 10/13 & 35/34 &    17/17 &  4/2 & 11/12 & 11/22 &  LC & st,emet & 937253 \\ 
           ...   &  ...  &  ... & ... & R & 20161009 & 22.7 &   3 & 450 &   VR &   0.4 &    11 &   ... &   ... &      ... &  ... &   ... &  ...  &  LC & st,emet & 937253 \\
\noalign{\smallskip} 
{\it 2016 WM1}   &   AP  &  G96 &  39 & P & 20161110 & 21.8 &   2 &  90 &   rg &   5.5 &    31 &  5/14 & 30/30 &    71/72 &  2/6 &   --- &  1/0  &  LC & st,met  & 924883 \\ % U growed! 
\noalign{\smallskip} 
{\bf 2017 EA}    &   AP  &  703 &   3 & P & 20170301 & 22.2 &   2 & 175 &    g &   1.2 &    13 &   1/2 & 49/49 &      1/1 &  3/3 &   3/0 &  4/0  &LC,DC& st,et   &1001439 \\
\noalign{\smallskip} 
2017 FB102       &   AP  &  F51 &  15 & P & 20170327 & 20.7 &   1 &  65 &    g &   8.1 &    33 &   6/6 & 25/25 &    59/59 &  6/6 &   5/5 &  5/6  &  LC & st,met  & 946657 \\
\noalign{\smallskip} 
{\it 2017 FV}    &   AP  &  G96 &  43 & P & 20170305 & 22.7 &   1 &  62 &    r &   0.5 &     2 & 16/28 & 34/33 &    25/25 &  6/4 &   --- &  3/0  &  LC & sl     & 937253 \\
           ...   &  ...  &  ... & ... & P & 20170306 & 22.6 &   1 & 127 &    g &   0.6 &     5 &   ... &   ... &      ... &  ... &   --- &  ...  &  LC & sl      & 937253 \\
\noalign{\smallskip} 
{\bf 2017 SE33}  &   AM  &  F51 &   9 & P & 20170920 & 20.1 &   1 & 112 &    g &  35.5 &   251 &   5/7 & 28/28 &    35/35 &  7/7 &   ?/0 &  2/0  &LC,DC&lt,met,F51&1020606\\ % removed before us from Sentry 
\noalign{\smallskip} 
{\bf 2017 SQ2}   &   AP  &  703 &  22 & P & 20170916 & 17.3 &   1 & 183 &    z &   9.0 &   104 & 13/15 & 23/25 &      8/8 &  6/1 &   1/0 &  3/0  &  LC & st,met  &1078705 \\ % removed before us both, Mega-Prec can't find it using both Desig and Orbit! 
           ...   &  ...  &  ... & ... & P & 20170917 & 18.3 &   1 & 200 &    g &   2.8 &    35 &   ... &   ... &      ... &  ... &   ... &  ...  &  LC & st,met  &1078705 \\ % Mega-Prec can't find it idem! 
\noalign{\smallskip} 
2017 UK3         &   AP  &  F51 &  12 & P & 20171020 & 20.7 &   1 &  90 &    i &   3.6 &    20 &   1/4 & 28/31 &    19/19 &  6/7 &   3/3 &  3/4  &  LC & st,met  &1078705 \\
\noalign{\smallskip} 
{\it 2017 UL1}   &   AP  &  G96 &  39 & R & 20171025 & 22.0 &   1 &  55 &    r &   1.5 &     5 &   5/9 & 42/43 &  111/112 &  8/8 &   --- &  1/0  &  LC & sl      &1081752 \\ % Sentry removed before us
           ...   &  ...  &  ... & ... & R & 20171025 & 22.0 &   1 &  63 &    r &   1.5 &     6 &   ... &   ... &      ... &  ... &   ... &  ...  &  LC & sl      &1081752 \\
\noalign{\smallskip} 
2017 UQ6         &   AT  &  G96 &  13 & P & 20171025 & 21.5 &   1 & 200 &    g &   4.5 &    57 & 17/19 & 27/26 &     8/8  &  7/7 &   2/1 &  -/2  &LC,DC& lt,met  &1022149 \\
\noalign{\smallskip} 
2017 VC14        &   AP  &  G96 &   7 & P & 20171111 & 22.5 &   1 &  80 &    g &   0.7 &     4 &   3/7 & 39/29 &     8/8  &  5/7 &   2/2 &  6/6  &  LC & sl      &1078706 \\
           ...   &  ...  &  ... & ... & P & 20171112 & 22.1 &   1 & 170 &    g &   1.1 &    12 &   ... &   ... &     ...  &  ... &   ... &  ...  &  LC & sle     &1078706 \\
\noalign{\smallskip} 
2018 BX5         &   AP  &  G96 &   5 & R & 20180126 & 21.4 &   2 &  35 &    i &   2.7 &     6 &   2/4 & 23/27 &     5/5  &  7/5 &  9/10 & 31/15 &  LC & sle     &1049192 \\
           ...   &  ...  &  ... & ... & R & 20180127 & 21.8 &   2 &  35 &    i &   1.8 &     4 &   ... &   ... &     ...  &  ... &   ... &  ...  &  LC & sle     &1049192 \\
\noalign{\smallskip} 
2018 DB          &   AP  &  G96 &  10 & R & 20180220 & 18.8 &   1 & 200 &    g & 133.8 &  1697 &   --- &   --- &     ---  &  --- &   --- &  ---  &  LC & vlt,nr  &    --- \\ % could not measure any end of trail, report only P (very fast rotator P~30)! 
\noalign{\smallskip} 
2018 EE9         &   AP  &  D29 &  18 & P & 20180308 & 17.7 &   1 & 150 &    r & 134.4 &  1273 & 14/18 & 18/17 &    24/24 &  8/8 & 21/26 & 40/32 &  LC & elt,met &1049204 \\ % Purple Mountain Observatory, XuYi Station; Mega-Prec can't find it using both Desig and Orbit! 
\noalign{\smallskip} 
2018 EZ2         &   AP  &  Q66 &  17 & P & 20180221 & 22.8 &   1 & 101 &    g &   0.6 &     4 & 66/85 & 37/37 &      8/8 &  4/4 &   1/1 &  2/2  &  LC & sl      &1084182 \\ % Siding Spring-Janess-G, JAXA japonezul?
           ...   &  ...  &  ... & ... & P & 20180221 & 22.8 &   1 &  40 &    r &   0.6 &     2 &   ... &   ... &      ... &  ... &   ... &  ...  &  LC & sl      &1084182 \\ 
\noalign{\smallskip} 
{\it 2018 FA4}   &   AP  &  G96 &  28 & P & 20180316 & 19.4 &   1 & 155 &    g &  28.3 &   277 & 26/28 & 25/25 &  278/278 &  6/4 &   --- &  1/0  &  LC & vlt,met &1057626 \\ 
\noalign{\smallskip} 
2018 FY2         &   AP  &  F51 &  24 & P & 20180315 & 20.6 &   1 & 109 &    r &  10.4 &    72 &   2/6 & 23/25 &  161/162 &  8/9 &  11/4 & 13/2  &DC,LC& lt,met  &1087522 \\
           ...   &  ...  &  ... & ... & P & 20180316 & 20.9 &   1 & 121 &    g &   7.7 &    59 &   ... &   ... &      ... &  ... &   ... &  ...  &DC,LC& lt,met  &1087522 \\
\noalign{\smallskip} 
2018 PZ21        &   AT  &  F51 &  16 & P & 20180811 & 20.1 &   1 &  56 &    g &  10.5 &    37 & 24/25 & 24/23 &     2/2  &  7/6 & 17/18 & 23/22 &LC,DC& st,oet  &1022150 \\
\noalign{\smallskip} 
2018 RF5         &   AM  &  G96 &  11 & P & 20180910 & 21.3 &   1 &  40 &    r &  30.4 &    77 &   7/9 & 26/26 &    12/12 &  5/5 &   3/3 &  3/3  &DC,LC& lt,met  &1024382 \\
           ...   &  ...  &  ... & ... & P & 20180911 & 20.2 &   3 &  90 &  rig &  25.3 &   144 &   ... &   ... &      ... &  ... &   ... &  ...  &DC,LC& lt,met  &1024382 \\
\noalign{\smallskip} 
2018 RQ4         &   AP  &  G96 &  14 & P & 20180907 & 21.6 &   1 &  95 &    g &   5.6 &    34 &  7/11 & 27/29 &    25/25 &  6/4 & 11/10 &  8/5  &LC,DC& st,met  &1024382 \\
           ...   &  ...  &  ... & ... & P & 20180907 & 21.6 &   1 &  46 &    r &   5.6 &    16 &   ... &   ... &      ... &  ... &   ... &  ...  &LC,DC& st,met  &1024382 \\
\noalign{\smallskip} 
{\bf 2018 RY1}   &   AT  &  G96 &  45 & P & 20180830 & 20.0 &   1 & 102 &    z &   6.0 &    39 & 26/47 & 30/29 &     8/8  &  6/6 &   ?/0 &  ?/0  &  LC &st,met,F51&1078709\\ % removed before us both
           ...   &  ...  &  ... & ... & P & 20180903 & 20.5 &   1 & 200 &    g &   4.9 &    62 &   ... &   ... &     ...  &  ... &   ... &  ...  &  LC &lt,met    &1078709\\
           ...   &  ...  &  ... & ... & P & 20180904 & 20.6 &   1 & 200 &    g &   4.6 &    58 &   ... &   ... &     ...  &  ... &   ... &  ...  &  LC &lt,met    &1078709\\
           ...   &  ...  &  ... & ... & P & 20180905 & 20.7 &   1 &  74 &    g &   4.4 &    21 &   ... &   ... &     ...  &  ... &   ... &  ...  &  LC &st,met    &1078709\\
           ...   &  ...  &  ... & ... & P & 20180906 & 20.9 &   1 &  56 &    g &   4.1 &    15 &   ... &   ... &     ...  &  ... &   ... &  ...  &  LC &st,met    &1078709\\
\noalign{\smallskip} 
2018 SD2         &   AT  &  T05 &   7 & P & 20180906 & 23.1 &   1 &  86 &    g &   0.6 &     3 &  2/17 & 46/52 &    11/11 &  6/6 & 48/33 &151/93 &DC,LC& sl      &1024382 \\
           ...   &  ...  &  ... & ... & P & 20180908 & 22.8 &   1 &  40 &    r &   0.6 &     2 &   ... &   ... &      ... &  ... &   ... &  ...  &DC,LC& sl      &1024382 \\
           ...   &  ...  &  ... & ... & P & 20180908 & 22.8 &   1 & 114 &    g &   0.6 &     4 &   ... &   ... &      ... &  ... &   ... &  ...  &DC,LC& sl      &1024382 \\
           ...   &  ...  &  ... & ... & P & 20180908 & 22.8 &   1 &  42 &    r &   0.6 &     2 &   ... &   ... &      ... &  ... &   ... &  ...  &DC,LC& sl      &1024382 \\
\noalign{\smallskip} 
{\it 2018 TB}    &   AM  &  T05 &  35 & P & 20180929 & 19.0 &   1 &  90 &    z &  21.4 &   122 & 48/51 & 48/48 &  110/110 &  4/3 &   --- &  1/1  &  LC & lt,met  &1078710 \\ % Sentry removed before us
\noalign{\smallskip} 
{\it 2019 DW1}   &   AP  &  G96 &  41 & P & 20190225 & 21.6 &   3 & 300 &    i &   4.4 &    83 &  9/12 & 27/29 &      7/7 &  8/7 &   2/0 &  ---  &DC,LC& lt,met  &1024384 \\
\noalign{\smallskip} 
2019 JH7         &   AP  &  G96 &   4 & P & 20190513 & 21.1 &   1 &  90 &    g &   2.2 &    12 &   2/3 & 88/59 &      4/4 &  3/3 &   2/2 &  2/2  &  LC & st,met  &1035768 \\
\noalign{\smallskip} 
{\bf 2019 MF3}   &   AP  &  F51 &  89 & P & 20190610 & 22.3 &   1 &  90 &    i &   2.2 &    12 & 12/31 & 22/22 &    27/27 &  6/2 &   ?/0 &  ?/0  &  LC &st,met,T14&1057633\\ % removed before us both
\noalign{\smallskip} 
\hline
%\end{tabular}
\end{longtable}
\end{center}
%\end{sidewaystable}
\end{landscape}
\end{tiny}

%______________________________________________________________

\end{document}